\documentclass[useAMS,usenatbib]{mn2e}
\usepackage{times}
\usepackage{graphicx}
\usepackage{float}   
\usepackage{epsfig}
\usepackage{chngcntr} 
\usepackage{caption} 
\usepackage{subcaption}
\usepackage{xspace}
\usepackage{booktabs} 
\usepackage{comment}  
\usepackage{amssymb} 
\usepackage{enumitem}   
\usepackage{rotating}
\usepackage[T1]{fontenc}
\usepackage{aecompl} 
\usepackage{tikz}

\newcommand{\msun}{\mathrm{M_{\odot}}}

\counterwithout{figure}{section}

\graphicspath{{/disk2/cxh/torus/paper/}} 

\title[Directly observing continuum emission from self-gravitating spiral waves]{Directly observing continuum emission from self-gravitating spiral waves}
\author[Cassandra Hall, Duncan Forgan, Ken Rice, Tim J. Harries, Pamela D. Klaassen and Beth Biller]{Cassandra Hall$^{1}$\thanks{Email: cxh@roe.ac.uk}, Duncan Forgan$^{2}$, Ken Rice$^{1}$,Tim J. Harries$^{3}$, Pamela D. Klaassen$^{4}$ and\newauthor Beth Biller$^{1}$\\
$^{1}$SUPA\thanks{Scottish Universities Physics Alliance}, Institute for Astronomy, University of Edinburgh, Blackford Hill, Edinburgh, EH9 3HJ, UK\\
$^{2}$SUPA\footnotemark[2], School of Physics \& Astronomy, University of St Andrews, North Haugh, St Andrews, KY16 9SS, UK\\
$^{3}$Department of Physics and Astronomy, University of Exeter, Stocker Road, Exeter EX4 4QL, UK\\
$^{4}$UK Astronomy Technology Center, Royal Observatory Edinburgh, Blackford Hill, Edinburgh, EH9 3HJ, UK\\
}
\begin{document}

\date{January 2016}

\pagerange{\pageref{firstpage}--\pageref{lastpage}} \pubyear{2015}

\maketitle

\label{firstpage}

\begin{abstract}
We use a simple, self-consistent, self-gravitating semi-analytic disc model to conduct an examination of the parameter space in which self-gravitating discs may exist. We then use Monte-Carlo radiative transfer to generate synthetic ALMA images of these self-gravitating discs to determine the subset of this parameter space in which they generate non-axisymmetric structure that is potentially detectable by ALMA.
Recently, several transition discs have been observed to have non-axisymmetric structure that extends out to large radii. It has been suggested that one possible origin of these asymmetries could be spiral density waves induced by disc self-gravity. We use our simple model to see if these discs exist in the region of parameter space where self-gravity could feasibly explain these spiral features. We find that for self-gravity to play a role in these systems typically requires a disc mass around an order of magnitude higher than the observed disc masses for the systems.
The spiral amplitudes produced by self-gravity in the local approximation are relatively weak when compared to amplitudes produced by tidal interactions, or spirals launched at Lindblad resonances due to embedded planets in the disc. As such, we ultimately caution against diagnosing spiral features as being due to self-gravity, unless the disc exists in the very narrow region of parameter space where the spiral wave amplitudes are large enough to produce detectable features, but not so large as to cause the disc to fragment.
\end{abstract}

\section{Introduction}
It is widely accepted that low-mass stars form through the collapse of cold, dense molecular cloud cores \citep{terebey1984,mckeeostricker2007}. These cores will typically contain some amount of angular momentum, meaning that all the mass cannot fall directly onto the central protostar; some must first pass through a protostellar accretion disc. In these discs, molecular viscosity alone does not exert large enough torques to redistribute angular momentum out to large radii, allowing mass to accrete onto the central protostar. However, instabilities that develop into the turbulent regime can produce considerable torques that can then drive mass transport.

If these discs are sufficiently massive then self-gravity could be  significant, and the gravitational instability (GI) could be the main angular momentum transport mechanism \citep{toomre1964,laughlin1994} during these early times. If GI is significant in these discs, then we would expect there to be non-axisymmetric structures, typically spiral density waves. 

Discs around very young stars are, however, heavily embedded in their cloud cores, making them difficult to observe at optical wavelengths \citep{dunhametal2014}. High resolution interferometric observations, in radio or sub-mm, are therefore required to resolve the disc. Currently, however, observations of this wavelength with a high enough resolution to resolve spiral arms are rare.

Here we examine the parameter space of self-gravitating protostellar discs to determine the range of accretion rates, disc masses and outer radii in which extended spiral features could be detected by ALMA. Previous studies which describe simulated ALMA observations of protostellar discs have tended to focus on reproducing the specific morphology of discs, using numerical methods such as Smoothed Particle Hydrodynamics (SPH), rather than an examination of the parameter space in which they are detectable \citep{cossinslodatotesti2010,rugeetal2013,douglasetal2013}.

\cite{dipierro2014} and \cite{dipierro2015} have shown, using simulated observations from SPH simulations, that non-axisymmetric structure in self-gravitating protostellar discs is detectable at a wide range of wavelengths using ALMA.  Our approach differs to that used in \cite{dipierro2014} and \cite{dipierro2015}, in that rather than deriving the physical disc structure from numerical simulations, where an artificial cooling law has been imposed, we use a self-consistent, analytic geometry coupled with 3D Monte Carlo radiative transfer (MCRT) to generate emission maps at typical ALMA frequencies. This geometry, unlike in SPH simulations by \cite{dipierro2014}, \cite{dipierro2015} and \cite{lodatorice2004}, is intended to be more realistic. We wish to note, however, that we do not include in our MCRT simulations the effect of dust trapping in the spiral arms noted by \cite{dipierro2015}. We suspect that including this would increase the detectability of the spiral structure, and we leave further investigation of this to future work.

Once we have our MCRT images, these are then used as input sky models to the ALMA simulator from the CASA software package (ver 4.3.0) to generate synthetic ALMA images. We stress that it is not our aim to match specific morphology, but rather to examine the conditions under which the over-density in spiral arms is sufficient so as to be detected. We then use this to investigate the region of parameter space (characterised by disc mass accretion rate and outer radius) in which GI-driven spiral density waves may be detectable by ALMA.

Currently, there are few, if any, observations which are strictly comparable with what we consider here. However, several transition discs have recently been observed to have non-axisymmetric structure that extends out to large radii. Bearing in mind the difficulty of finding strictly comparable samples, we take three of these transition discs as test cases, and apply our simple geometry to them. We aim to determine if these discs exist in the parameter space where self-gravity could be a feasible explanation for their spiral structure. We do not, however, generate synthetic images of these cases, since these test cases are imaged in NIR and scattered light, which is quite different from continuum mm emission. 

The paper is laid out as follows: The model is described in Section \ref{sec:model}, with the 2D and 3D structure given in Sections \ref{subsec:2D} and \ref{subsec:3D} respectively. Section \ref{subsec:TORUS} describes our Monte Carlo radiative transfer method, and Section \ref{subsec:ALMA} describes the generation of synthetic observations. Our results are given in Section \ref{sec:results}. We give the surface density profiles achieved from our model in Section \ref{subsec:basicresults}, and compare our models to scenarios where a fixed cooling law is imposed in Section \ref{subsec:betaresults}. We discuss our general results from our synthetic imaging in Section \ref{subsec:resultsALMA}, and apply our analytical geometry to test cases in Sections \ref{subsubsec:MWC758}, \ref{subsubsec:SAO206462} and \ref{subsubsec:HD142527}. We draw conclusions in Section \ref{sec:conclusions}.
\section{Model}
\label{sec:model}
\subsection{Outline}
\label{subsec:Outline}
\begin{figure}
\begin{center}
\includegraphics[scale=0.55]{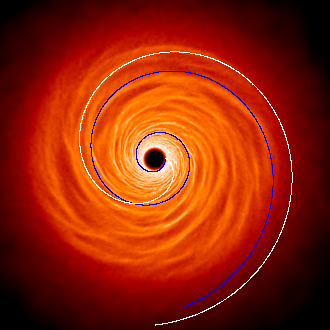}\\
\caption{This is a simulation image of a self-gravitating disc that has reached a state of quasi-equilibrium, with parameterised cooling such that $\beta=9$. Two spirals have been plotted. The blue line is a logarithmic spiral, the white line is an Archimedal spiral. The logarithmic spiral appears to fit the spiral arms of the disc the best, although it is fairly arbitrary since Equation (\ref{eq:spiral}) has two free parameters.\label{fig:spirallines}}
\end{center}
\end{figure}

Here we use an existing 1D model to examine the parameter space of self-gravitating discs initially developed by \citet{clarke2009} (see also \citealt{ricearmitage2009,forganrice2013}). We develop it to include 2D and 3D structure, fitting spirals of the shape typically found in Smoothed Particle Hydrodynamics (SPH) simulations (given in Figure \ref{fig:spirallines}).
\subsection{Radial Geometry}
\label{subsec:2D}
We expect a self-gravitating protostellar system to settle into a state of quasi-steady thermal equilibrium \citep{paczynski1978,gammie2001,ricearmitage2009},  with a constant accretion rate $\dot{M}$ given by
\begin{equation}
\label{eq:mdot}
\dot{M}=\frac{3\pi\alpha c_s^2\Sigma}{\Omega},
\end{equation}
where $c_s$ is the sound speed, $\Sigma$ is the surface density, $\Omega$ is the angular frequency (since we are considering Keplerian rotation, the epicyclic frequency is simply $\Omega$) and $\alpha$ is the dimensionless shear stress, composed of both hydrodynamic (i.e. Reynolds stress) and gravitational parts. Remarkably, this can be expressed both simply and analytically as \citep{gammie2001}
\begin{equation}
\label{eq:alphaeff}
\alpha_{\mathrm{grav}}=\frac{4}{9}\frac{1}{\gamma(\gamma-1)\beta},
\end{equation}
where $\gamma$ is the ratio of specific heats and $\beta$ is a dimensionless constant which parameterises cooling, and is given by 
\begin{equation}
\label{eq:beta}
\beta =t_c \Omega,
\end{equation}
where $t_c$ is the cooling time. The scale height $H$ is given by
\begin{equation}
\label{eq:H}
H=\frac{c_s}{\Omega}
\end{equation}
and the midplane density is given by
\begin{equation}
\rho_0=\frac{\Sigma}{2H}.
\end{equation}
For a chosen $\dot{M}$ and outer radius $R$, we iterate our code until the surface density $\Sigma$ produces the accretion rate we are attempting to match.  A disc is susceptible to gravitational instability if the \citet{toomre1964} parameter $Q \approx 1.5 - 1.7$ \citep{durisenetal2007}. For our purposes, we impose the following condition at all radii in the disc,
\begin{equation}
\label{eq:Q}
Q=\frac{c_s\Omega}{\pi \mathrm{G}\Sigma}=2,
\end{equation}
where G is the gravitational constant. With the chosen accretion rate and outer radius fixed, we then find $\Sigma$ by guessing an overly large value and iterating downwards until $\dot{M}$ in Equation (\ref{eq:mdot}) matches our chosen $\dot{M}$. For a given value of $\Sigma$, we obtain the local sound speed by rearranging Equation (\ref{eq:Q}). This allows the calculation of the local scale height, and hence the midplane density. An equation of state table is then used to determine the opacity $\kappa$ from this density and sound speed, and then the optical depth $\tau$ is estimated as $\tau=\kappa\Sigma$ using Rosseland mean opacities from \citet{belllin1994}. The cooling rate is then \citep{hubeny1990}:
\begin{equation}
\label{eq:coolrate}
\Lambda = \frac{8\mathrm{\mathrm{\sigma}}T^4}{3\tau},
\end{equation}
where $T$ is the midplane temperature and $\mathrm{\sigma}$ is the Stefan-Boltzmann constant. If some source of external irradiation is present, this is modified to
\begin{equation}
\label{eq:coolrateirr}
\Lambda = \frac{8\mathrm{\mathrm{\sigma}}(T^4-T_{\mathrm{irr}}^4)}{3\tau},
\end{equation}
where $T_{\mathrm{irr}}$ is the temperature set by the external irradiation. The cooling time $t_c$ in both cases is then the thermal energy per unit area divided by the cooling rate:
\begin{equation}
\label{eq:cooltime}
t_c=\frac{1}{\Lambda}\frac{c_s^2\Sigma}{\gamma(\gamma-1)}.
\end{equation}
Since in a quasi-steady state, energy dissipation in a disc is dominated by self-gravity \citep{lodatorice2004}, and assuming local thermodynamic equilibrium, the rate per unit time, per unit area at which the kinetic energy of rotation of the disc is dissipated into heat by viscosity (the dissipation rate) is equal to the cooling rate, given by Equation \ref{eq:coolrate}. This dissipation rate is 
\begin{equation}
\label{eq:dissrate}
\mathcal{D}(R)=\frac{1}{2}\nu_{\mathrm{vis}}\Sigma \Bigg(R\frac{\partial\Omega}{\partial R}\Bigg)^2,
\end{equation}
where $\nu_{\mathrm{vis}}$ is kinematic viscosity, and is expressed as \citep{shakurasunyaev1973}
\begin{equation}
\label{eq:viscosity}
\nu_{\mathrm{vis}} = \frac{\alpha c_s^2}{\Omega}.
\end{equation}
This allows $\alpha$ to absorb the uncertainties of the pseudo-viscous properties of the disc. Since in a steady state, the cooling rate matches the dissipation rate, we can equate Equations (\ref{eq:coolrate}), or (\ref{eq:cooltime}) and (\ref{eq:dissrate}), and then use Equation (\ref{eq:viscosity}) to directly determine $\alpha$. Finally, since the accretion rate in a quasi-steady disc is given by Equation (\ref{eq:mdot}), we can check that our calculated $\dot{M}$ is within some tolerance of our imposed $\dot{M}$. If not, we reduce $\Sigma$ and repeat this until $\dot{M}$ is within some tolerance of our chosen $\dot{M}$. Therefore, the surface density is determined iteratively at every radius and only needs to be integrated to determine the enclosed mass of the disc. 
\begin{figure}
\begin{center}
\includegraphics[scale=0.4]{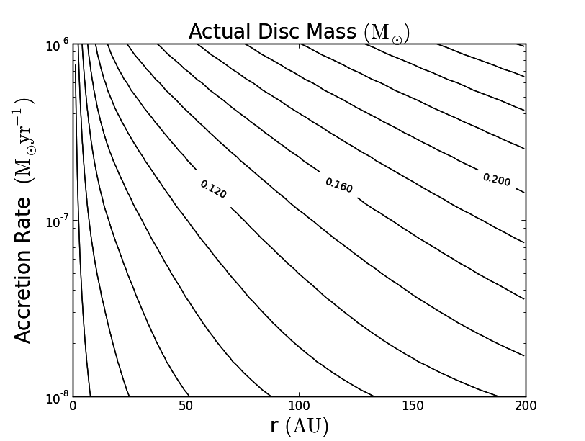}
\end{center}
\caption{2D contour lines of actual disc mass (around a 1 M$_{\odot}$ star) as a function of accretion rate and radius for self-gravitating discs with no external irradiation. 
\label{fig:disccontour}}
\end{figure}
The parameter space of our model is shown in Figure \ref{fig:disccontour}. For every accretion rate, and radius at that accretion rate, there is a corresponding disc mass. 
\subsection{3D Structure}
\label{subsec:3D}
The above method only examines the radial parameter space. To examine structures such as spiral arms, the model must be developed to manage azimuthal asymmetry. Additionally, since we later compute global 3D radiative transfer calculations, we also include a vertical density profile. We begin by defining a Cartesian grid onto which surface density as a function of radius is mapped. This produces a 2D disc which is azimuthally uniform at each radius. It has been shown there is a relationship between the amplitude of the density perturbations and the strength of the cooling such that \citep{cossinslodatoclarke2009}
\begin{equation}
\bigg(\frac{\delta\Sigma}{\Sigma}\bigg)^2=\frac{2}{\epsilon\beta}\frac{1}{\gamma(\gamma -1)}\bigg(\frac{1}{\mathcal{M}\mathcal{\widetilde{M}}}\bigg),
\end{equation}
where $\epsilon$ is a dimensionless proportionality factor known as the heating factor, and the radial phase and Doppler-shifted phase Mach numbers are $\mathcal{M}=|v_{\mathrm{p}}|/c_s$ and $\mathcal{\widetilde{M}}=|\tilde{v}_{\mathrm{p}}|/c_s$. \citet{cossinslodatoclarke2009} find an empirical relationship such that 
\begin{equation}
\frac{\langle\delta\Sigma\rangle}{\langle\Sigma\rangle}\approx \beta^{-\frac{1}{2}}.
\end{equation}
Since there is a relationship between $\alpha$ and $\beta$ given by Equation (\ref{eq:alphaeff}), we impose a spiral perturbation %
dictated by our $\alpha_{\mathrm{grav}}$ such that \citep{riceetal2011}
\begin{equation}
\label{eq:alphadsigma}
\frac{\langle\delta\Sigma\rangle}{\langle\Sigma\rangle}\approx \alpha_{\mathrm{grav}}^{1/2},
\end{equation}  
where $\delta\Sigma/\Sigma$ is the fractional over-density. This is imposed sinusoidally so that
\begin{equation}
\delta\Sigma(\phi) = -\langle\delta\Sigma\rangle\cos(m\phi),
\end{equation} 
where $m$ is the azimuthal wavenumber selected by the user. For self-gravitating discs with mass ratios similar to what we'll be considering, this is typically high, so we select 10. We have assumed that the maximum over-density is equal to the average over-density, with $\phi$ the phase difference between the azimuthal location of the spiral $\theta_{\mathrm{spiral}}$ and the azimuthal position $\theta_{\mathrm{x,y}}$ in the disc:
\begin{equation}
\phi=\theta_{\mathrm{spiral}}-\theta_{\mathrm{(x,y)}}.
\end{equation}
The maximum over-density then occurs when the (x,y) position is coincident with the position of the spiral. The negative sign in Equation (\ref{eq:alphadsigma}) simply forces the maxima to occur coincident with the position of the spiral. A logarithmic spiral of the form 
\begin{equation}
\label{eq:spiral}
\theta_{\mathrm{spiral}} = \frac{1}{b}\log\Bigg(\frac{r}{a}\Bigg),
\end{equation}
was used, where $a$ and $b$ are constants (in this case $a=13.5$ and $b=0.38$), as this most closely matches the shape of spirals seen in self-gravitating discs from simulation data (see Figure \ref{fig:spirallines}). This is somewhat arbitrary, since tweaking parameters can give a close fit with an Archimedean spiral. However, it is not our intention to exactly match morphology.

To create accurate skymodels in terms of brightness to put into the ALMA simulator, the vertical density profile must be carefully considered, since the total intensity at the surface of the disc is dependent upon the amount of emission and absorption which has occurred between the surface and the midplane. We calculate the density as a function of $z$ using the expression for density in a self-gravitating disc (see e.g. \cite{spitzer1942} for a full derivation)
\begin{equation}
\label{eq:rhoznow}
\rho(z)=\rho_0\Bigg[\frac{1}{\mathrm{cosh}^2\big(\frac{z}{H_{\mathrm{sg}}}\big)}\Bigg],
\end{equation}
where the self gravitating scale height $H_{\mathrm{sg}}$ is given by
\begin{equation}
\label{eq:Hsg}
H_{\mathrm{sg}}=\frac{c_s^2}{\pi\mathrm{G}\Sigma}.
\end{equation}
It is worth noting that $H_{sg}$ is approximately equal to $H$ in Equation (\ref{eq:H}) since 
\begin{equation}
\frac{H_{\mathrm{sg}}}{H}=\frac{c_s\Omega}{\pi\mathrm{G}\Sigma}=Q=2.
\end{equation}
%
\subsection{Radiative and Molecular Line Transfer Code: TORUS}
\label{subsec:TORUS}
The TORUS radiation transport code \citep{harriesetal2004,kurosawaetal2004,haworthetal2015} determines radiative equilibrium in a dusty medium using the Monte Carlo (MC) photon packet method first described by \cite{lucy1999}. Temperatures, densities and dust properties are stored on a
three-dimensional adaptive mesh, refined in such a way that large
gradients in opacity are well resolved. Here we use a cylindrical
adaptive mesh, in which the cells are sectors of hollow
cylinders. When additional resolution is required a cell may be
subdivided into four (or eight) children by splitting the radial
extent and height of the cell into two, and (optionally) by splitting
the azimuthal extent of the sector into two.

We use a simple thermal model for the disc geometry, and then calculate a radiative-equilibrium model using TORUS. The radiation equilibrium calculation is iterative and full details
are given in the above references. Briefly, the radiation field of
the protostar is modelled using $N$ photon packets which are allowed to
propagate through the grid, undergoing a random walk of scatterings, or
absorptions and re-emissions, until they escape the computational
domain, at which point estimates are made of the absorption rate in
each cell. The dust temperatures in the grid are then calculated on
a cell-by-cell basis by assuming radiative equilibrium, and the next
iteration of the photon loop is performed (using the updated dust
temperatures). Once the temperatures have converged it is possible to
calculate spectral energy distributions and continuum images for
arbitrary viewing angles using the MC method. The TORUS code has been
extensively benchmarked and shows good agreement with other
independently developed radiative transfer codes \citep{pinteetal2009}.

For our main radiative transfer results (in Section \ref{subsec:resultsALMA}) we assume typical values for a pre-main-sequence star, with central source mass of $M_{*} = \mathrm{M}_{\odot}$, $R_*= 2.325$ $R_{\odot}$ and $T_{\mathrm{eff}} = 4350$ K. The dust in our model consists of \citet{drainelee} silicates, with a grain size distribution of
\begin{equation}
\label{eq:grainsize}
n(a)\propto a^{-q} \hspace{10mm} \mathrm{for} \hspace{10mm}a_{\mathrm{min}}<a<a_{\mathrm{max}},
\end{equation}
where $a_{\mathrm{min}}$ and $a_{\mathrm{max}}$ are the minimum and maximum size of the dust grains ($0.1$ $\mu$m and $1.0$ $\mu$m respectively), and the power-law exponent $q$ is taken to be $q_{\mathrm{ism}}=3.5$ \citep{grainsize}. The dust density is 1\% of the gas density.     
%
%
%
\subsection{The ALMA Simulator}
\label{subsec:ALMA}
The emission maps generated by TORUS are used as inputs to the ALMA simulator built into CASA (ver 4.3) \citep{casa}.  Disc sizes and fluxes are scaled to a distance of $\sim 140$ pc (i.e. in Taurus), and we show a set of comparison images at $\sim 50$ pc (i.e. in TW Hydrae).

 Multiple simulations were conducted varying the array size, and therefore imaging resolution and sensitivity, to ensure the optimum balance between resolution and sensitivity. Built-in noise sources such as atmospheric transmissions were included in the simulations run at varying precipitable water vapour (PWV) levels.  Typical PWV levels appropriate for observing in the different ALMA bands were used, specifically: 2.784, 1.262, 1.262 and 0.472 mm for simulated observations at 220, 345, 460 and 680 GHz, respectively. These are consistent with those used, for instance, in \citet{dipierro2014}.
\begin{figure*}
\begin{tabular}{ccc}
\hspace{7 mm}$\dot{M}=2.8\times 10^{-7}$ M$_{\odot}$ yr$^{-1}$ & \hspace{7 mm}$\dot{M}=3.2\times 10^{-7}$ M$_{\odot}$ yr$^{-1}$ & $\dot{M}=3.6\times 10^{-7}$ M$_{\odot}$ yr$^{-1}$\\
\rotatebox{90}{\hspace{20 mm}\textbf{No Irradiation}}
\includegraphics[trim = 70 15 80 40, clip,width=0.31\textwidth]{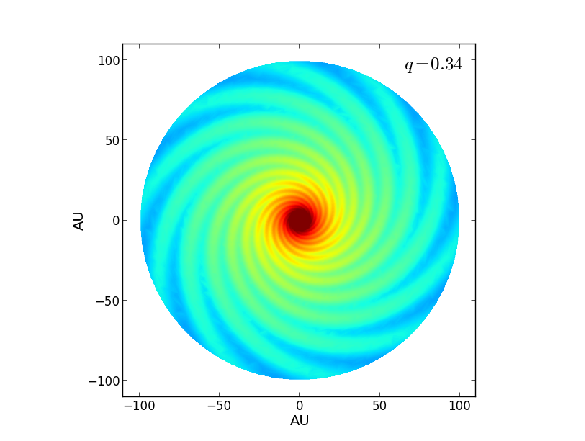} &\includegraphics[trim = 70 15 80 40, clip,width=0.31\textwidth]{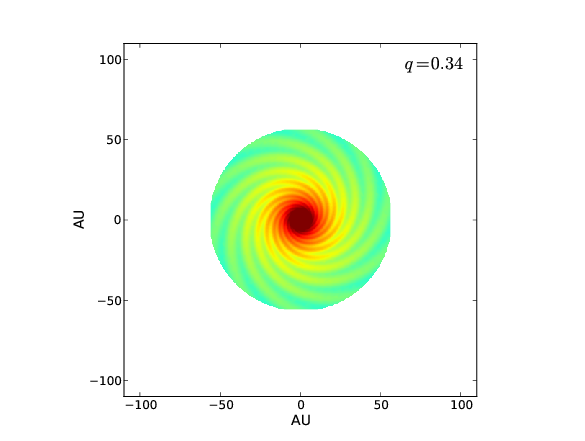}&\includegraphics[trim = 30 15 40 40, clip,width=0.3675\textwidth]{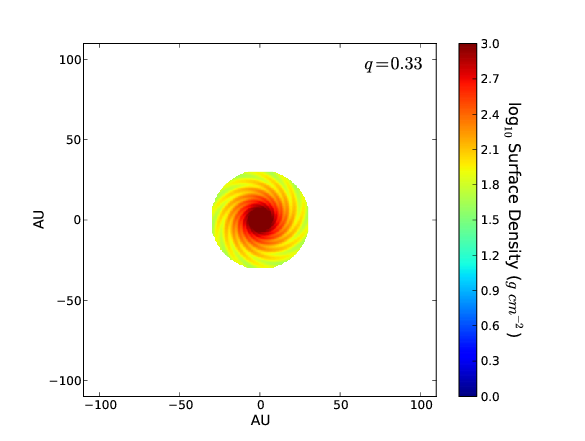}\\
\rotatebox{90}{\hspace{13 mm}\textbf{Irradiation at 10 K}}
\includegraphics[trim = 70 15 80 40, clip,width=0.31\textwidth]{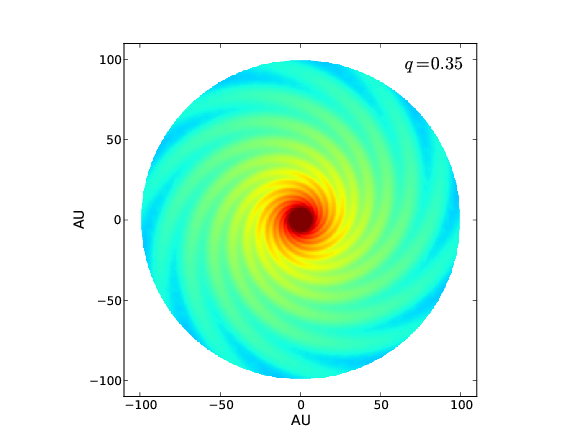} &\includegraphics[trim = 70 15 80 40, clip,width=0.31\textwidth]{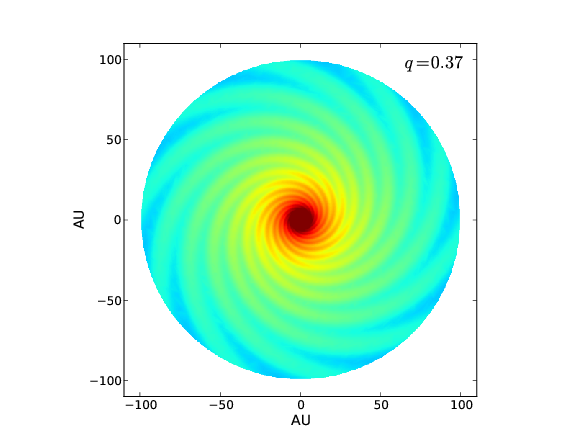}&\includegraphics[trim = 30 15 40 40, clip,width=0.3675\textwidth]{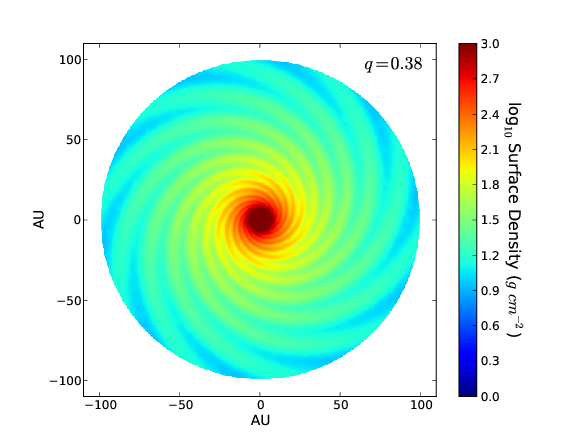}\\
\rotatebox{90}{\hspace{13 mm}\textbf{Irradiation at 30 K}}
\includegraphics[trim = 70 15 80 40, clip,width=0.31\textwidth]{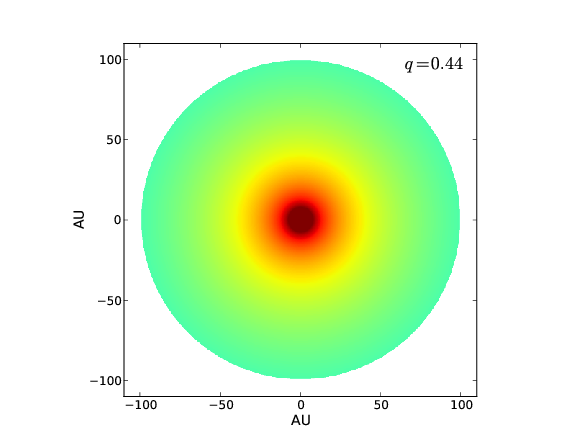} &\includegraphics[trim = 70 15 80 40, clip,width=0.31\textwidth]{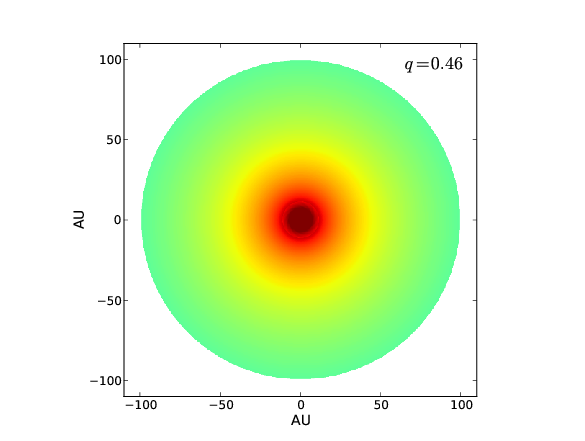}&\includegraphics[trim = 30 15 40 40, clip,width=0.3675\textwidth]{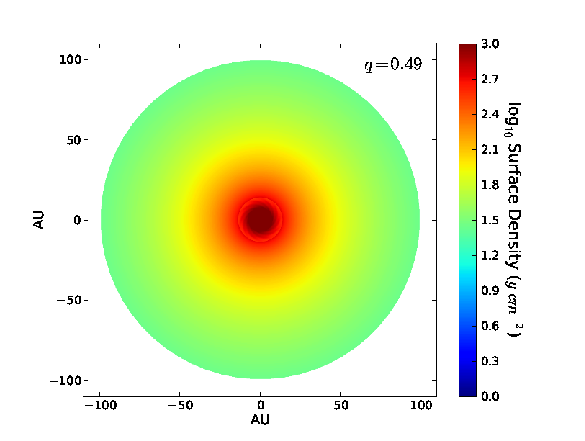}\\
\end{tabular}
    \caption{Logarithmic surface density maps of discs with accretion rate increasing from left to right. Top row: For a 1$\msun$ central star and no external irradiation, the maximum accretion rate a 100 AU disc can sustain without the outer regions fragmenting is $\dot{M}= 2.8\times 10^{-7}\mathrm{M}_{\odot}$ $\mathrm{yr}^{-1}$. Any higher, and the disc truncates in the outer regions as it becomes unstable to collapse and begins to fragment. Middle row: same three accretion rates, but with external irradiation at 10K added. Although the discs look similar, they vary very slightly in mass. The irradiation is just enough to prevent truncation, without removing spiral features. Bottom row: same discs but with irradiation at 30K added. These discs are also prevented from fragmentation in the outer regions, however the irradiation is also sufficient to remove spiral features completely.}\label{fig:truncation}
\end{figure*}
\section{Results}
\label{sec:results}
Our results section is broken into four parts. We begin with the basic results from our analytic geometry (Section \ref{subsec:basicresults}), showing that for a given central star mass and disc outer radius, there are regions of parameter space in which the disc cannot exist in a quasi-steady, self-gravitating state. We then show that such discs may exist in these regions of parameter space if they are irradiated with some external source, but a moderate amount of irradiation will remove spiral features.

In Section \ref{subsec:betaresults}, we compare the synthetic ALMA image results of our analytical model to image results which used SPH geometries. We show that the assumptions made in SPH simulations probably cause larger contrast in the inner regions of discs. We do, however, reproduce the basic results of \cite{dipierro2014} using a constant $\beta$ approach.

In Section \ref{subsec:resultsALMA}, we present results showing the conditions required to directly observe self-gravitating spiral structures with ALMA. We entered a range of accretion rates, and found that even at the maximum accretion rate a 100 AU disc can sustain without fragmenting, non-axisymmetry due to disc self-gravity is just detectable at 680 GHz at a distance of 140 pc (i.e. in Taurus). At 50 pc (i.e. in TW Hydrae), it is significantly easier to detect spiral structure, and the features are discernible at 220 GHz. 

Finally, we make \emph{only} parameter space comparisons with three observed systems in Section \ref{subsec:comparison}. We show that for all three systems, it seems unlikely that spiral features which have recently been imaged in the disc are due to disc self-gravity, unless the disc mass has been significantly underestimated.

\subsection{Basic Model}
\label{subsec:basicresults}
We find that for a given radius and host star mass, with no external irradiation considered, there is a maximum accretion rate that any self-gravitating, quasi-steady disc can sustain. Above this accretion rate, the disc begins to truncate as the outer parts become susceptible to fragmentation \citep{rafikov2005,forganrice2011}. This is illustrated in the top panel of Figure \ref{fig:truncation}, which shows logarithmic surface density plots of three discs with increasing accretion rate from left to right. Note that to decrease the disc radius to a quarter of its original size only requires an increase in accretion rate of around 30\%.

When more mass is added to the disc, the sound speed (and subsequently temperature from our equation of state) increases since in our model it is set by Equation (\ref{eq:Q}). In the cool outer parts of the disc, the Rosseland mean opacity is related to temperature by $\kappa\propto T^2$ \citep{whitworthetal2010}. Therefore our cooling rate now has a dependence $\Lambda\propto T^2$. This increased local cooling rate causes a decrease in local cooling time. In order to maintain the disc in a quasi-steady state, this local radiative cooling must be balanced by viscous shock heating from the spiral arms, therefore the local $\alpha_{\mathrm{grav}}$ increases to redress the balance in the disc. 

However, this quasi-steady, self-gravitating torque saturates at around $\alpha\sim 0.1$ \citep{gammie2001,rice2005} and we expect the disc to fragment, producing bound objects. Since this region of parameter space is not what we are interested in, we set the surface density here to 0.

However, a small amount of irradiation, say at 10K, can change the surface density profile of a disc. How this changes depends upon the accretion rate of the disc, and therefore how close $\dot{M}$ was initially to the fragmentation/truncation boundary in the absence of irradiation. 

The middle panels of Figure \ref{fig:truncation} show discs with an accretion rate that was initially at the fragmentation boundary before irradiation at 10K was added. In this case, the irradiation prevents truncation whilst preserving spiral structure. The disc mass varies very slightly, and this is consistent with previous examinations of disc mass under irradiation \citep{forganricejeans2013} since the accretion rate is only changing slightly, rather than by an order of magnitude. We therefore find that a small amount of external irradiation can prevent fragmentation, whilst preserving the spiral features in the disc.
\begin{figure}
\begin{center}
\includegraphics[scale=0.45]{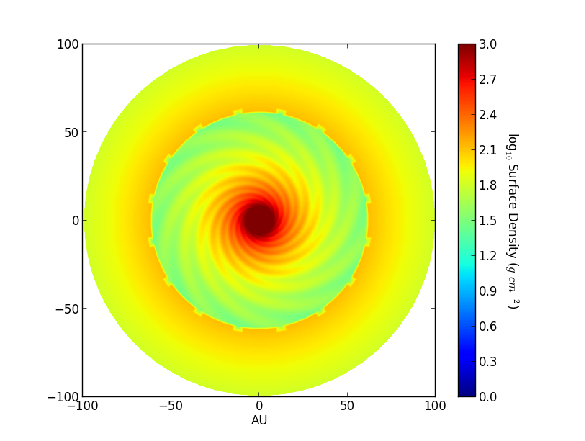}\\
\caption{Disc with $\dot{M}=2.0\times 10^{-7}$ M$_{\odot}$ yr$^{-1}$, 33\% lower than the maximum it can sustain without truncating. It is irradiated at 10 K, and this small amount of external irradiation has altered the surface density structure of the outer parts of the disc, as the equilibrium disc structure is now more massive for a given accretion rate and radius. The ``spikey'' features on the boundary between the outer and inner parts of the disc are a numerical artefact due to a small change in surface density from the resolution limit of the grid, and is more pronounced in log space.\label{fig:MWCT10K}}
\end{center}
\end{figure}

On the other hand, we do find that spiral features in a self-gravitating disc of any accretion rate are erased by a moderate amount of external irradiation, say at 30K. This is demonstrated by the bottom panels in Figure \ref{fig:truncation}, which shows that for a disc with a radius of 100 AU, irradiation at a temperature of 30K can erase non-axisymmetric structure from the disc. It is unsurprising that applying sufficient external irradiation to our systems wipes out spiral structure, as external irradiation has been found to have a stabilising effect on marginally unstable discs \citep{krattermurrayclay2011,riceetal2011,forganricejeans2013}. When external irradiation of temperature $T_{\mathrm{irr}}$ is present, the local cooling rate in the disc is reduced and is given by Equation (\ref{eq:coolrateirr}). This decrease in cooling rate causes the cooling time $t_c=\beta/\Omega$ to increase, which also causes local $\beta$ to increase. If the disc is in thermal equilibrium, we then parameterise the viscous shock heating required to balance this cooling as $\alpha_{\mathrm{grav}}$ given by Equation (\ref{eq:alphaeff}), where $\alpha_{\mathrm{grav}}$ can be thought of as an effective gravitational stress. It is easy to see that for an increased $\beta$, a smaller $\alpha$ is required to redress the balance. The $\alpha$ is then able to stay below the torque saturation limit $\alpha_{\mathrm{crit}}\sim 0.1$ \citep{gammie2001,rice2005}, and the disc is able to stay in a quasi-steady self-gravitating state out to larger radii. Although in this manner, the disc is able to stay in a quasi-steady, self-gravitating state, the strength of the spiral amplitudes decrease. In our calculations, the $\alpha$ is composed of two parts,
\begin{equation}
\alpha=\alpha_{\mathrm{grav}}+\alpha_{\mathrm{visc}},
\end{equation}
where $\alpha_{\mathrm{visc}}$ is the viscous component of the stress due to magnetorotational instability (MRI), which we set to $\alpha_{\mathrm{visc}}=0.01$ (see e.g. \citealt{kratteretal2008}). Since the perturbation strength of the spiral amplitudes is given by
\begin{equation}
\label{eq:dsigmaNOW}
\frac{\langle\delta\Sigma\rangle}{\langle\Sigma\rangle}=\alpha^{\frac{1}{2}}_{\mathrm{grav}},
\end{equation}
when $\alpha\rightarrow 0$ then $\frac{\langle\delta\Sigma\rangle}{\langle\Sigma\rangle}\rightarrow 0$. This happens when the midplane temperature $T$ in Equation (\ref{eq:coolrateirr}) tends to $T_{\mathrm{irr}}^4$. In this case, cooling rate $\Lambda\rightarrow 0$, so cooling time $t_c\rightarrow \infty$ and $\beta\rightarrow\infty$. This means $\alpha_{\mathrm{grav}}\rightarrow 0$ by Equation (\ref{eq:alphaeff}), and $\frac{\langle\delta\Sigma\rangle}{\langle\Sigma\rangle}\rightarrow 0$. In this limit, cooling is balanced by irradiation and $\alpha_{\mathrm{visc}}$, so $\alpha_{\mathrm{grav}}$ is no longer needed for thermal equilibrium. Essentially, increasing temperature provides extra pressure support against gravitational collapse. In our geometry, however, we do not consider infalling mass from a nascent cloud. This could potentially reverse our result of small amounts of irradiation halting fragmentation, as this infalling mass causes a positive rate of change of local mass, decreasing the Jeans mass and potentially encouraging fragmentation, provided that $Q$ remains constant. 

Lowering the accretion rate (and therefore disc mass), changes the effect of a given amount of external irradiation. Figure \ref{fig:MWCT10K} illustrates this. It is a disc with an accretion rate of $\dot{M}=2.0\times 10^{-7}$ M$_{\odot}$ yr$^{-1}$, approximately a third lower than in Figure \ref{fig:truncation}. In this case (Figure \ref{fig:MWCT10K}) external irradiation has added additional mass to the outer parts of the disc, as the equilibrium disc structure is now more massive for a given accretion rate and radius, although in this case it does not meet the criteria required for it to fragment (which requires that the local $\alpha \gtrsim 0.1$). In this disc, the external irradiation is maintaining $Q\sim 2$ with little dissipation of the gravitational instability. In reality, we should expect $Q$ to increase beyond the marginal limit of self-gravity in the outer parts of such discs.

We should bear in mind that this is a simple model in which no infall is considered. Since, in general, adding mass to the outer regions of a self-gravitating disc encourages fragmentation \citep{kratteretal2010a,vorobyovbasu2010,krattermurrayclay2011,forganrice2012}, it seems that the Jeans criterion for fragmentation may be satisfied at relatively high accretion rates (of order $\sim 10^{-7}$ M$_{\odot}$ yr$^{-1}$) in the presence of sufficiently small ($\sim 10$ K) irradiation.

\begin{figure}
\begin{center}
\includegraphics[scale=0.45]{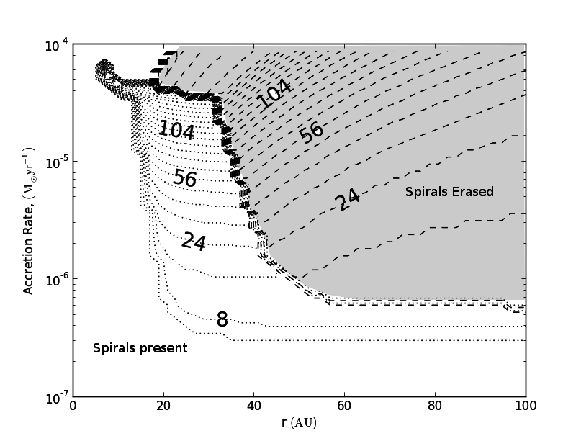}
\caption{Contour plot showing the \emph{minimum} temperature in (K) of external irradiation required to halt fragmentation as a function of accretion rate and radius. Shaded region with dashed contours shows where both fragmentation is suppressed and spiral features are erased. The unshaded region with dotted contours shows where added irradiation prevents fragmentation but preserves spiral structure.\label{fig:tempcontours}}
\end{center}
\end{figure}

Figure \ref{fig:tempcontours} is a contour plot of the amount of external irradiation required to prevent fragmentation of the disc as a function of accretion rate and radius. Below $\sim 3.2\times 10^{-7}$ M$_{\odot}$ yr$^{-1}$, in our local viscous approximation, any disc will be able to regulate itself against collapse. Above this, differing amounts of irradiation can either prevent fragmentation \emph{and} totally remove spiral features from the disc, or prevent fragmentation whilst allowing spiral features to exist. As accretion rate increases, higher temperatures are required to prevent the disc from fragmenting; additionally, beyond 60 AU the determining factor in whether spirals are present or not is temperature, rather than a combination of temperature and accretion rate.
\subsection{Comparison with imposed constant $\beta$}
\label{subsec:betaresults}
\begin{figure}
\begin{tabular}{cc}
Realistic $\alpha$ & Constant $\beta$\\
\rotatebox{90}{\hspace{13 mm}220 GHz}
\includegraphics[trim=82 3 8 8,clip,width=0.48\linewidth]{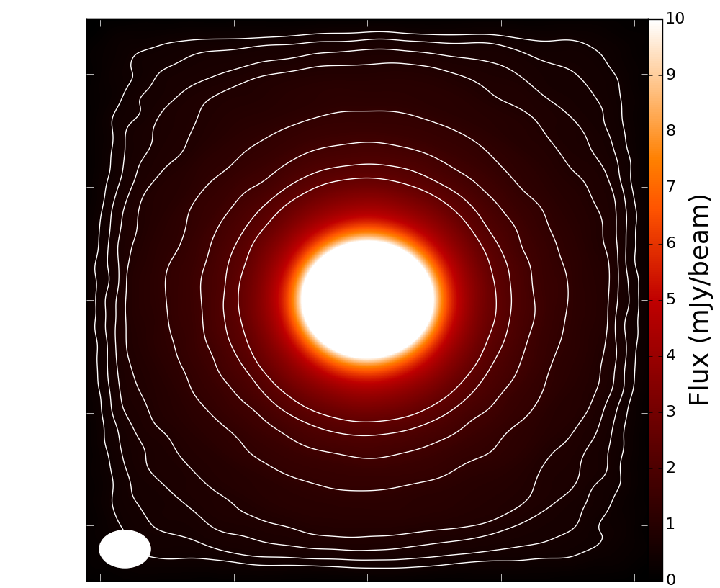} & \includegraphics[trim=82 3 8 8,clip,width=0.48\linewidth]{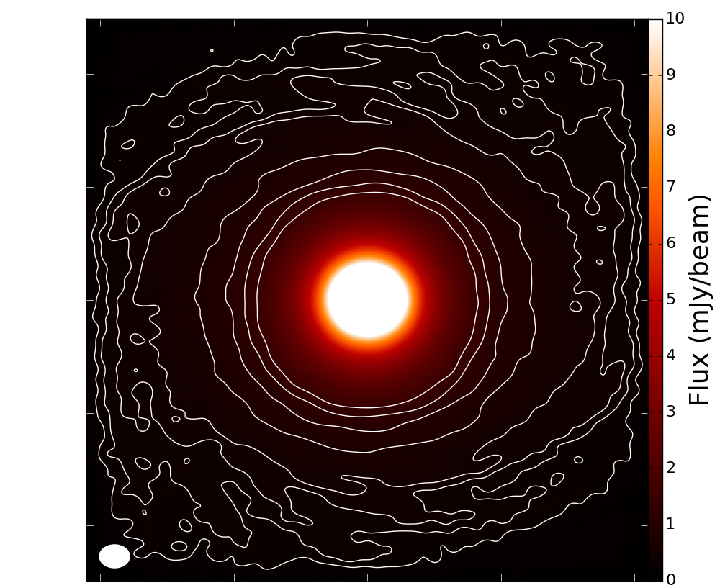}\\
\rotatebox{90}{\hspace{9 mm}680 GHz}
\includegraphics[trim=82 3 8 8 ,clip,width=0.48\linewidth]{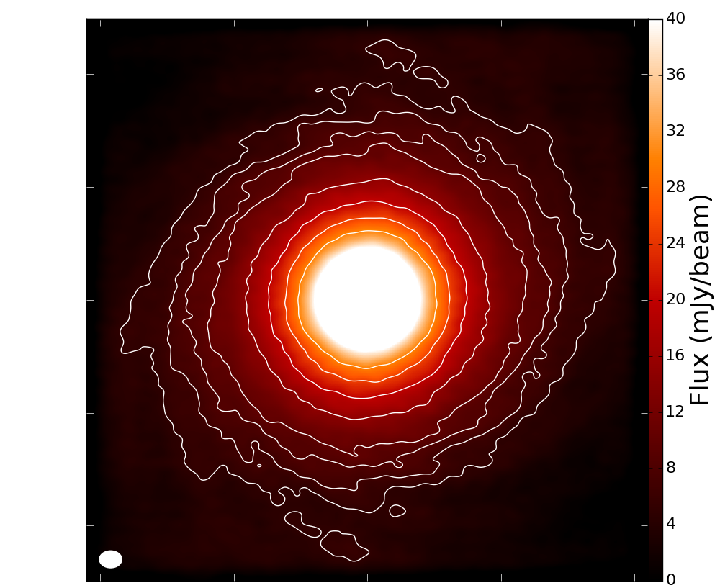} & \includegraphics[trim=82 3 8 8 ,clip,width=0.48\linewidth]{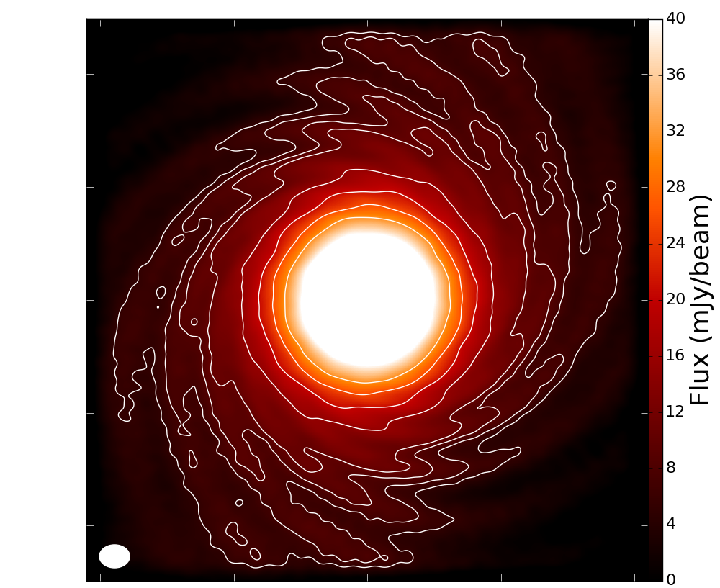}\\
\multicolumn{2}{|c|}{\includegraphics[trim=20 15 45 40,clip,width=1.0\linewidth]{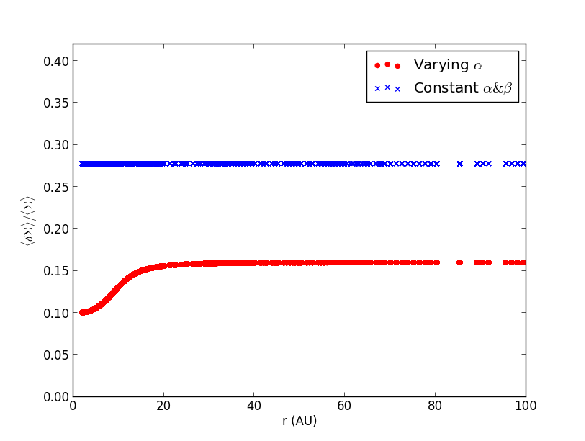}}
\end{tabular}
\caption{Top 4 panels: Synthesised ALMA images for two $R=100$ AU discs with different geometries. Both discs have a central star mass of 1 M$_{\odot}$ and central star radius 1 R$_{\odot}$, with $T_{\mathrm{source}}=6000K$, imaged at a distance of 50 pc for clarity. Beam size which gives the best compromise resolution and sensitivity is selected for each disc, details are given in Table \ref{table:ALMA}. Left column has realistic $\alpha$, whereas right column has in imposed constant $\beta$ (and therefore $\alpha$) value. Contours are at 3, 4, 5, 6, 9, 12, 15 \& 18 $\times$ the RMS in each image. Each image has been scaled to best show the spiral features (if present). Disc mass is the same for both ($q=0.21$), $\dot{M}=5\times 10^{-8}$ M$_{\odot}$ yr$^{-1}$ for realistic $\alpha$ and  $\dot{M}=2.8\times 10^{-7}$ M$_{\odot}$ yr$^{-1}$ for constant $\beta$ - and, hence, constant $\alpha$. In both cases, the spiral arms are more clear for $\beta=6$. Bottom panel: Spiral amplitude strength vs radius for the two discs. Red dots are ``realistic'' $\alpha$ and blue crosses are fixed $\alpha$($\beta$).}
  \label{fig:B6} 
\end{figure}
In our model, unlike in SPH simulations by \cite{dipierro2014} and \cite{lodatorice2004}, we do not artificially impose a constant gravitational $\alpha_{\mathrm{grav}}$ by imposing a constant $\beta$. Doing so implies that the spiral amplitude strength $\langle\delta\Sigma\rangle/\langle\Sigma\rangle \approx \alpha_{\mathrm{grav}}^{1/2}$ is constant at all radii. In our model we take a more realistic approach and allow $\alpha$ to vary, requiring that our disc remain marginally unstable to non-axisymmetric perturbations (i.e. $Q=2$), and that the cooling rate at each radius depends on the local conditions. In such a quasi-steady, self-gravitating disc we expect $\alpha$ to increase with increasing radius \citep{ricearmitage2009}. 

The basic consequence of this is that the assumption of a constant $\beta$ means that there will clearly be regions where the $\alpha$ values will be quite different to what would be the case if more realistic assumptions were used.  In particular, the $\alpha$ values would probably be significantly larger in the inner regions, which will cause a greater contrast than would be the case were more realistic assumptions used.  Figure \ref{fig:B6} compares two systems, both with the same total mass, but one determining $\alpha$ realistically (left hand panel) and the other assuming that $\beta$ (and hence $\alpha$) is constant, fixed at $\beta = 6$.  The reason we choose the same disc mass, rather than the same accretion rate, is that the total disc flux $F_{\mathrm{D}}$ is related to total disc mass $M_{\mathrm{D}}$ by
\begin{equation}
F_D=\frac{2\kappa(\nu)\mathrm{k_B}\nu^2T_{\mathrm{dust}}M_{\mathrm{disc}}}{D^2c^2},
\end{equation}
where $\kappa(\nu)$ is opacity, $\mathrm{k_B}$ is the Boltzmann constant, $c$ is the speed of light and $D$ is the distance to the object. The disc mass is the same in both cases ($q=0.2$), but holding $\beta$ fixed alters the geometry so the equilibrium accretion rate for a given disc mass differs. Therefore $\dot{M}=5\times 10^{-8}$ M$_{\odot}$ yr$^{-1}$ for realistic $\alpha$ and  $\dot{M}=2.8\times 10^{-7}$ M$_{\odot}$ yr$^{-1}$ for constant $\beta$ yield the same total disc mass.  

In the fixed $\beta$ (and therefore fixed $\alpha$) scenario, the strength of the spiral amplitudes are larger throughout the disc and so they are easily detectable to $3\sigma$ confidence level even at $\nu=220$ GHz. At $\nu=680$ GHz, the interarm regions are a low enough temperature to broach the Wien limit, thus increasing the contrast ratio between the spiral arms and the inter-arm regions. 

The first thing that we should stress is that using our semi-analytic model with fixed beta, we are able to reproduce the basic results of \cite{dipierro2014}. We find that extended structure is detectable to 3 $\times$ the RMS noise, and that the fluxes agree to the same order of magnitude in both the realistic $\alpha$ case and imposed constant $\beta$ (hence $\alpha$) case, giving us some confidence that our general method is reasonable. 

We have, however, assumed our dust grains are dynamically well coupled to the gas such that $\rho_{\mathrm{dust}}$ is linearly proportional to $\rho_{\mathrm{gas}}$. \cite{dipierro2015} showed that dust trapping occurs in the spiral arms in self-gravitating discs, however this predominantly occurs for particles with sizes of a centimetre or more \citep{riceetal2004}. Therefore, these overdensities are best probed at frequencies that are not considered here ($\sim 10$ GHz), and so should not change our results.

However, what Figure \ref{fig:B6} also shows is that in the fixed $\beta$ scenario, the strength of the spiral amplitudes are larger throughout the disc than in the realistic $\alpha$ case.  At $\nu = 680$ GHz, this translates to more clearly discernible spiral structure throughout the disc. We could, of course, choose a larger $\beta$ value, or increase the mass in the realistic alpha case, so that there were at least regions where the amplitudes were comparable. However, it is clear that even in such cases, the constant $\beta$ assumption would produce unrealistic amplitudes in the inner parts of these discs. 

For the rest of this paper we allow the effective gravitational stress $\alpha_{\mathrm{grav}}$ to vary as it would do in a ``realistic'' disc. 
\subsection{ALMA Images}
\label{subsec:resultsALMA}
\begin{table*}
\begin{tabular}{cccccccc}
\hline
Figure  & $\nu$ (GHz)& Distance (pc) &Beam size (asec)   & RMS ($\mu$Jy beam$^{-1}$)& Contours ($\times$ RMS) &Integration time (s) & PWV (mm)\\
\hline
\hline
\ref{fig:B6} & 220 &50  &  0.23 $\times$ 0.17 &       141      & 3,4,5,6,9,12,18        &  1800                & 2.784         \\
             & 220 &50  &  0.13 $\times$ 0.10 &        82      & 3,4,5,6,9,12,18        &  1800                & 2.784         \\
             & 680 &50  &  0.13 $\times$ 0.11 &      1674      & 3,4,5,6,9,12,18        &  7200                & 0.472         \\
             & 680 &50  &  0.13 $\times$ 0.10 &        1732    &  3,4,5,6,9,12,18       &  1800                & 0.472         \\    
\hline
\ref{fig:ALMAlower:a} & 220 & 140  & 0.0706 $\times$ 0.0601 &       20     & 3,5,7,9,12,15,18          &  7200       & 2.784         \\
\ref{fig:ALMAlower:c} & 220 & 140 & 0.0706 $\times$ 0.0601 &        22    & 3,5,7,9,12,15,18          &  7200       & 2.784         \\
\ref{fig:ALMAlower:e} & 220 & 140 & 0.0706 $\times$ 0.0601   &    20      & 3,5,7,9,12,15,18          &  7200       & 2.784         \\
\ref{fig:ALMAlower:b} & 680 & 140  & 0.0435 $\times$ 0.0331   &    196     & 4,7,10,13,15,20          &  7200       & 0.472         \\ 
\ref{fig:ALMAlower:d} & 680 & 140 & 0.0435 $\times$ 0.0331 &      269   & 3,5,7,9,12,15,18          &  7200       & 0.472         \\
\ref{fig:ALMAlower:f} & 680 & 140 & 0.0435 $\times$ 0.0331   &       253   & 3,5,7,9,12,15,18          &  7200       & 0.472         \\
\hline
\ref{fig:distancecomparison} & 220 & 140   &  0.0706 $\times$ 0.0601 &       20      & 3,5,7,9,12,15,18        &  7200                & 2.784         \\
             & 220 & 50   &  0.135 $\times$ 0.102 &        60   & 4,7,10,13,15,18        &  7200                & 2.784         \\
             & 680 & 140  &  0.0435 $\times$ 0.0331 &      196      & 4,7,10,13,15,20        &  7200                & 0.472         \\
             & 680 & 50  &  0.0732 $\times$ 0.0539 &    720    & 4,7,10,13,15,20       &  7200                & 0.472         \\  
\hline
\end{tabular}
\caption{Image details for figures shown in this work. We detail frequency of the synthesised observation, whether a realistic or constant $\alpha$ was used in the image, the accretion rate, disc mass, the size of the beam, the noise of the image, the integration time used and precipitable water vapour (PWV) value.}

\label{table:ALMA}
\end{table*}
\begin{table*}
\begin{tabular}{cccccc}
\hline
Figure  & $\alpha$ type &$M_*$ (M$_{\odot}$) &$\dot{M}$ (M$_{\odot}$ yr$^{-1}$)& $q$ ($M_{\mathrm{D}}/M_*$)\\
\hline
\hline
\ref{fig:B6}    & Realistic &1& $5\times 10^{-8}$ & 0.21   \\
                & Constant  &1& $2.8\times 10^{-7}$ & 0.21  \\
                & Realistic &1& $5\times 10^{-8}$  & 0.21  \\
                & Constant  &1& $2.8\times 10^{-7}$ & 0.21  \\
\hline
\ref{fig:ALMAlower:a}   & Realistic&1 & $2.8\times 10^{-7}$ & 0.34 \\
\ref{fig:ALMAlower:c}   & ''        &1& $1.0\times 10^{-7}$& 0.25  \\
\ref{fig:ALMAlower:e}   & ''        & 1&$5.0\times 10^{-8}$ & 0.21 \\
\ref{fig:ALMAlower:b}   & ''        &1& $2.8\times 10^{-7}$ & 0.34 \\
\ref{fig:ALMAlower:d}   & ''        &1& $1.0\times 10^{-7}$ & 0.25 \\
\ref{fig:ALMAlower:f}   & ''        &1& $5.0\times 10^{-8}$ & 0.21 \\
\hline
N/A       &           &              &               \\
MWC 758     & Realistic & 2&$2.0\times 10^{-7}$ & 0.25 \\
SAO 206462  & ''        & 1.7&$5.37 \times 10^{-9}$ & 0.1 \\
HD 142527   & ''        & 2&$6.92 \times 10^{-8}$& 0.75  \\
\hline
\end{tabular}
\caption{Physical parameters of the discs used to create synthesis images in this work. Columns are figure number, whether a realistic or constant $\alpha$ was used, host star mass, accretion rate and disc-to-star mass ratio}

\label{table:parameters}
\end{table*}
\begin{figure*} 
\begin{center}  
  \begin{subfigure}[b]{0.33\linewidth}
 \captionsetup{width=.9\linewidth}%

     \rotatebox{90}{\hspace{19 mm}220 GHz} \includegraphics[trim=85 15 0 0, clip, width=0.96\linewidth]{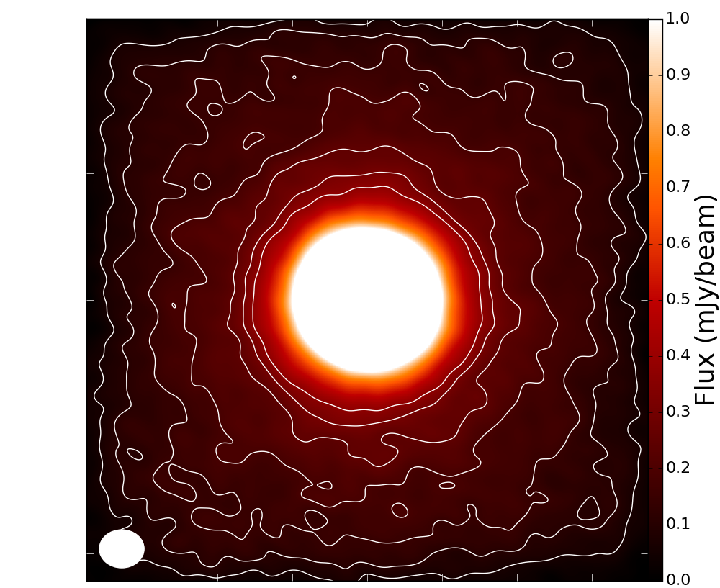}

    \caption{$2.8\times 10^{-7}\mathrm{M}_{\odot}$ $\mathrm{yr}^{-1}$,\\ RMS = $20$ $\mu\mathrm{Jy}$ beam$^{-1}$. } 
    \label{fig:ALMAlower:a} 

  \end{subfigure}
  \begin{subfigure}[b]{0.33\linewidth}
 \captionsetup{width=.9\linewidth}%
    \centering
 
    \includegraphics[trim =85 15 0 0, clip,width=0.96\linewidth]{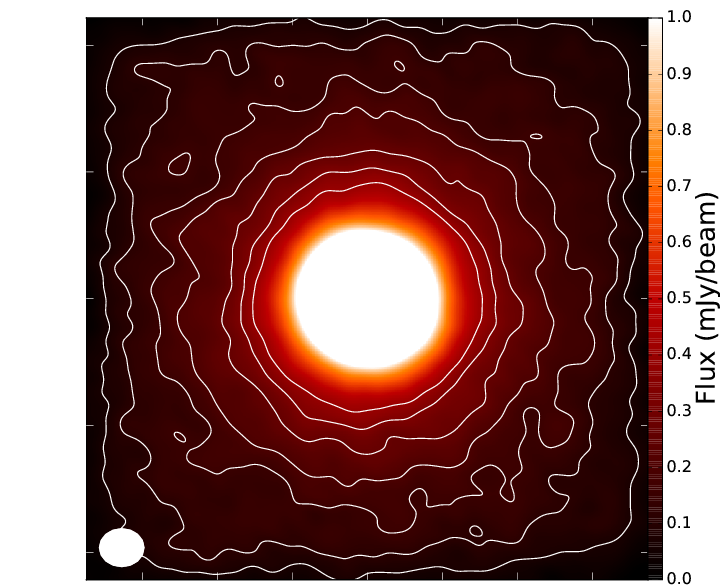}
    \caption{$1\times 10^{-7}\mathrm{M}_{\odot}$ $\mathrm{yr}^{-1}$, \\RMS = $22$ $\mu\mathrm{Jy}$ beam$^{-1}$.} 
    \label{fig:ALMAlower:c} 
  \end{subfigure}
  \begin{subfigure}[b]{0.33\linewidth}
 \captionsetup{width=.9\linewidth}%
    \centering
    \includegraphics[trim =85 15 0 0, clip,width=0.96\linewidth]{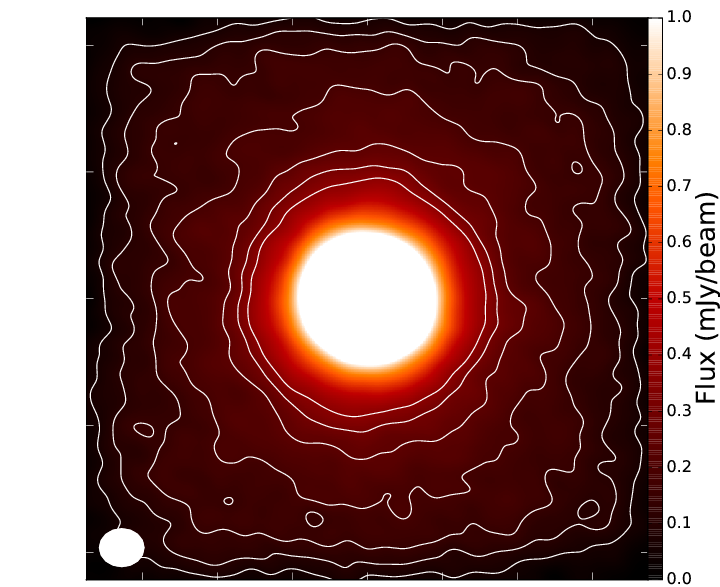}
    \caption{$5\times 10^{-8}\mathrm{M}_{\odot}$ $\mathrm{yr}^{-1}$,\\ RMS = $20$ $\mu\mathrm{Jy}$ beam$^{-1}$.} 
    \label{fig:ALMAlower:e} 
  \end{subfigure}\\
  \begin{subfigure}[b]{0.33\linewidth}
 \captionsetup{width=.9\linewidth}%
 \rotatebox{90}{\hspace{19 mm}680 GHz} \includegraphics[trim =85 15 0 0, clip,width=0.96\linewidth]{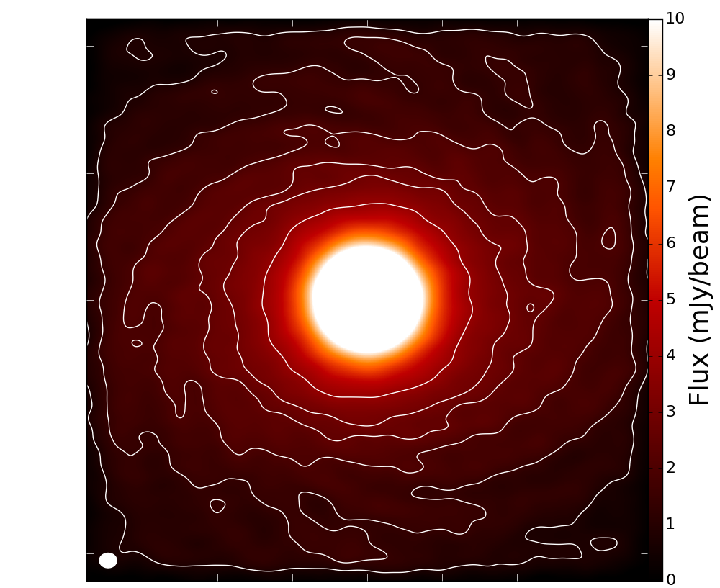}
    \caption{$2.8\times 10^{-7}\mathrm{M}_{\odot}$ $\mathrm{yr}^{-1}$,\\ RMS = $196$ $\mu\mathrm{Jy}$ beam$^{-1}$.} 
    \label{fig:ALMAlower:b} 
  \end{subfigure} 
  \begin{subfigure}[b]{0.33\linewidth}
 \captionsetup{width=.9\linewidth}%
    \centering
    \includegraphics[trim =85 15 0 0, clip,width=0.96\linewidth]{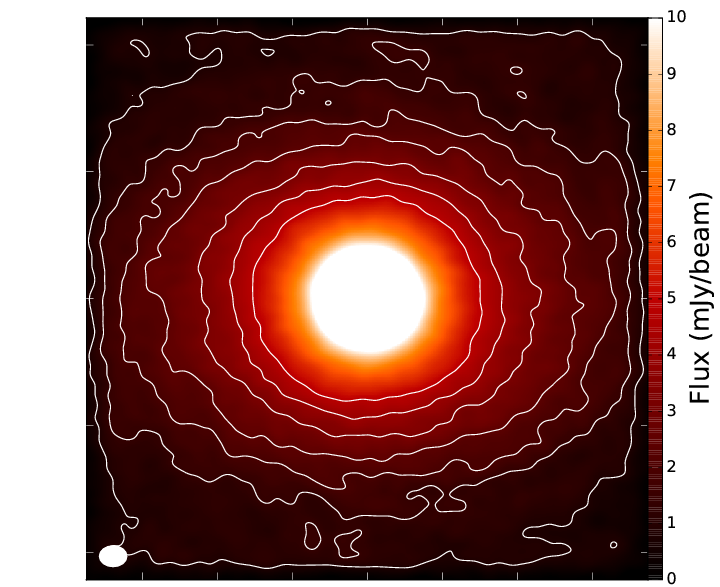}
    \caption{$1\times 10^{-7}\mathrm{M}_{\odot}$ $\mathrm{yr}^{-1}$,\\ RMS = $269$ $\mu\mathrm{Jy}$ beam$^{-1}$.} 
    \label{fig:ALMAlower:d} 
  \end{subfigure} 
  \begin{subfigure}[b]{0.33\linewidth}
 \captionsetup{width=.9\linewidth}%
    \centering
    \includegraphics[trim =85 15 0 0, clip,width=0.96\linewidth]{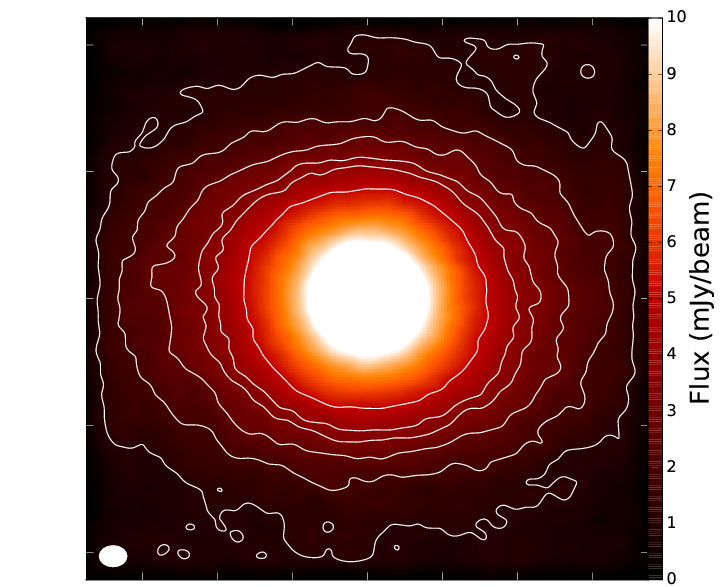}
    \caption{$5\times 10^{-8}\mathrm{M}_{\odot}$ $\mathrm{yr}^{-1}$,\\ RMS = $253$ $\mu\mathrm{Jy}$ beam$^{-1}$. } 
    \label{fig:ALMAlower:f} 
  \end{subfigure} 
\end{center}
  \caption{Synthesised ALMA images for $R=100$ AU discs with accretion rate decreasing from left to right. Top row is at 220 GHz, bottom row is at 680 GHz. All discs are imaged at 140 pc. Left column is the maximum accretion rate it can sustain without fragmenting, $\dot{M}= 2.8\times 10^{-7}\mathrm{M}_{\odot}$ $\mathrm{yr}^{-1}$. Contours are at multiples of the RMS given in Table \ref{table:ALMA}. Beam size is in bottom left corner, beam details are given in Table \ref{table:ALMA}. Geometry details are given in Table \ref{table:parameters}. Below $\dot{M}= 5.0\times 10^{-8}\mathrm{M}_{\odot}$ $\mathrm{yr}^{-1}$, at 220 GHz flux from the disc is low enough that thermal noise dominates, so the central region (inner 20 AU) is detectable, but not the extended non-axisymmetric structure. At 680 GHz, asymmetry is noticeable, but the spiral arms are not clearly defined.}
  \label{fig:ALMAlower} 
\end{figure*}
In this section we present results illustrating the conditions under which we may be able to directly observe self-gravitating spiral density structures using ALMA.  We assume that all discs have outer radii of 100 AU and follow the procedure described in Section \ref{sec:model} to determine the amplitude of the spiral density waves, the continuum emission from the disc, and what we would then expect to be observed by ALMA.

Figure \ref{fig:ALMAlower} shows six synthetic ALMA images of three 100AU discs around a 1.0$\msun$, 2.325 R$_{\odot}$ star, with a surface temperature of $T=4350$ K, and with three different accretion rates. All discs are imaged at a distance of 140 pc. Each disc is observed at 220 GHz (top row) and 680 GHz (bottom row). Observing parameters are given in Table \ref{table:ALMA}, and physical parameters are given in Table \ref{table:parameters}.

From left to right, the accretion rates are $\dot{M} = 2.8\times 10^{-7}$ M$_{\odot}$ yr$^{-1}$, $\dot{M} = 1.0\times 10^{-7}$ M$_{\odot}$ yr$^{-1}$ and $\dot{M} = 5.0\times 10^{-8}$ M$_{\odot}$ yr$^{-1}$. Figures \ref{fig:ALMAlower:a} and \ref{fig:ALMAlower:b} depict a disc with the maximum accretion rate $\dot{M} = 2.8\times 10^{-7}$ M$_{\odot}$ yr$^{-1}$ a disc of outer radius $R=100$ AU, with no external irradiation, can sustain without fragmenting. Non-axisymmetric structure is only visible at the higher frequency of 680 GHz. We wish to stress at this point that this is the absolute maximum accretion rate that this disc, with this particular set of parameters and no external irradiation, can sustain.

For a sufficiently high accretion rate, and therefore disc-to-star mass ratio $q$, we reproduce the results of \citet{dipierro2014} of increasing contrast with increasing frequency, as the Planck law in the interarm regions falls into the Wien limit due to the low temperature. 

Additionally, Figure \ref{fig:ALMAlower} shows that as accretion rate (and therefore disc mass) is decreased, the central part of the disc remains detectable to at least the 3$\sigma$ level, but the spiral structure is simply not detectable with ALMA at 220 GHz (or longer wavelengths). Non-axisymmetry is, however, noticeable at higher frequency (shorter wavelengths) since the lower temperatures in the inter-arm regions means the Planck law is in the Wien limit for these frequencies, reducing the intensity of emission.
\begin{figure}
\begin{tabular}{cc}

\textbf{140 pc} & \textbf{50 pc}\\
\rotatebox{90}{\hspace{12 mm}\textbf{220 GHz}}
\includegraphics[trim=82 3 5 8,clip,width=0.48\linewidth]{review220GHzm28e-7.png} &\includegraphics[trim=82 30 8 8,clip,width=0.48\linewidth]{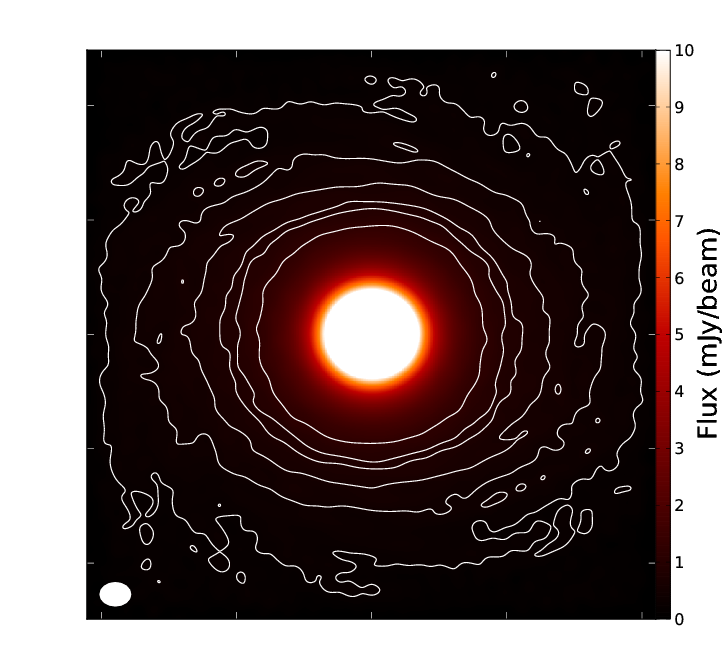} \\
\rotatebox{90}{\hspace{12 mm}\textbf{680 GHz}}
\includegraphics[trim=82 3 8 8,clip,width=0.48\linewidth]{review680GHzm28e-7.png}&
\includegraphics[trim=82 3 8 8 ,clip,width=0.48\linewidth]{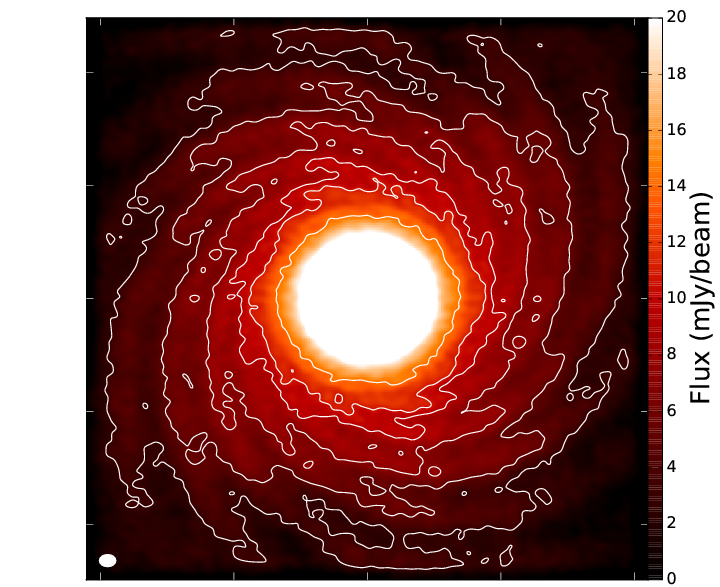}
\end{tabular}
\caption{All discs are an outer radius of 100 AU, imaged at 220 GHz and 680 GHz, at a distance of 50pc and 140 pc. M$_* = 1 $M$_{\odot}$, R$_* = 2.325$ R$_{\odot}$ and $T_{*} = 4350$ K. Right column shows the disc imaged at a distance of 50 pc, as in the TW Hydrae region. Left column shows discs imaged at a distance of 140 pc, as in the Taurus region. Contours are at intervals of the RMS, given in Table \ref{table:ALMA}, beam sizes are also in Table \ref{table:ALMA}.}
  \label{fig:distancecomparison} 
\end{figure}

The images in Figure \ref{fig:ALMAlower} show the difficulty faced when determining the presence of GI in a protostellar disc. Since a distance of 140 pc corresponds to a typical star forming region (Taurus), these are not the most optimistic results. Figure \ref{fig:distancecomparison} shows the same disc, with an accretion rate of $\dot{M}=2.8\times 10^{-7}$ M$_{\odot}$ yr$^{-1}$, imaged at 140 pc (220 GHz and 680 GHz), and 50 pc (at 220 GHz and 680 GHz). At a closer distance, it is significantly easier to detect spiral structure in the disc, as shown in the images at 50 pc. At a distance of 50 pc, the spiral structure in the outer part of the disc is also detectable at 220 GHz, whereas it is not detectable at 140 pc. 

It becomes more difficult to detect spiral structure with decreasing accretion rate and disc mass. Since the strength of our spiral amplitudes are determined by $\delta\Sigma/\Sigma \approx \alpha^{1/2}$, when the accretion rate is lowered, so are the spiral amplitudes. The conclusion we can draw from this examination of parameter space is that in quasi-steady, self-gravitating discs, for any given disc radius and host star mass, there exists a narrow range of accretion rates in which the outer part of the disc does not begin to fragment, but for which the spiral structure is detectable. Additionally, even if the disc is within this parameter space, the distance to the object and the frequency of the observations may also determine the likelihood of spiral structure being detected. 

With this in mind, it is prudent to caution against diagnosing directly imaged non-axisymmetric structure as due to disc self-gravity, unless the disc in question is sufficiently massive. That our discs of lower disc mass/accretion rate fail to produce detectable spiral features appears to conflict with the result found by \cite{dipierro2014}. However, as previously mentioned, our model allows both the cooling time and $\alpha$ to vary locally, so the relative strengths of our perturbations are much less in the outer part of the disc than they would be in an SPH simulation where $\beta$ is fixed at some relatively low value.
\subsection{Comparison With Observed Systems}
\label{subsec:comparison}
Currently, there aren't any observations that are directly comparable to what we present here. There are, however, some systems with spiral features typically observed in NIR scattered light. We consider the properties of such systems and compare them to what we suggest would be required if self-gravity is to be the source of the spiral features.
\subsubsection{MWC 758}
\label{subsubsec:MWC758}
The transition disc around Herbig A5 star MWC 758 is located in the edge of the Taurus star forming region at a distance of 279$^{+94}_{-48}$ pc \citep{vanleeuwen2007}. It is $3.5\pm 2$ Myr old \citep{meeusetal2012}and has a stellar mass of 2 M$_{\odot}$ \citep{isellaetal2010}. The disc mass and radial extent are approximated from sub-millimetre observations as $10^{-2}$ M$_{\odot}$ and $\sim 100$ AU respectively \citep{andrewsetal2011}. The accretion rate is estimated as somewhere between $2\times 10^{-7}$ M$_{\mathrm{\odot}}$ yr$^{-1}$ \citep{isellaetal2008} and $1\times 10^{-8}$ M$_{\mathrm{\odot}}$ yr$^{-1}$ \citep{andrewsetal2011}.

The first near-IR (NIR) scattered light images, clearly showing the discovery of spiral arms, were given in \citet{gradyetal2013}, obtained using Subaru and 1.1 $\mu$m \emph{Hubble Space Telescope}/NICMOS data. The parameterised fit of the spiral arms was performed by \citet{gradyetal2013} following \citet{mutoetal2012}. It is possible that such spirals are launched by a perturbing body, if so it would require a mass of $\sim 5$ M$_{\mathrm{J}}$, which is consistent with continued accretion onto the central star.  

\citet{marinoetal2015} combine VLA Ka and ALMA maps to show that the disc is clearly non-axisymmetric. The disc is fit with a steady state vortex solution to explain the spiral arms. The authors suggest that the compact emission in VLA Ka data is consistent with an accreting companion object such as a forming planet, which could also be responsible for the spiral arms imaged in scattered light. The companion planet scenario is consistent with simulations conducted by \citet{dongzhu2015}.

Similarly, MWC 758 was imaged in scattered light by \citet{benistyetal2015}, using VLT/SPHERE to achieve a higher resolution than previously achieved. The spirals arms were again modelled using density wave theory, with two planetary companions launching the spiral arms. Although the spirals are interpreted as being due to planetary companions, other mechanisms, such as GI, can launch spiral waves with low $m$ modes that are capable of matching these observed features, as shown by \citet{donghallrice2015}. The measured disc mass ($10^{-2}$ M$_{\odot}$) of MWC 758 is probably too low to trigger gravitational instabilities (see e.g. \citealt{gammie2001}), however, as discussed in our introduction, there are large uncertainties in the ratio of dust to gas and there is evidence that T-Tauri disc masses have been systematically under-estimated. 

Observations have revealed a complex morphology of the disk of MWC 758. To understand the origin of these spiral features, both modelling and high resolution images in the sub-mm with ALMA is needed. Scattered light traces the surface variations in a disc (a valid assumption for vertical isothermal hydrostatic equilibrium), whilst to probe structures near the midplane it is preferable to use longer wavelengths with high spatial resolution. 

In this work, we model MWC 758 as if it is self-gravity that is responsible for these spirals, to see if it is indeed the likely origin of these features. We simply assess whether it exists in the parameter space required for self-gravity to exist.

\begin{figure}
\includegraphics[width=\linewidth]{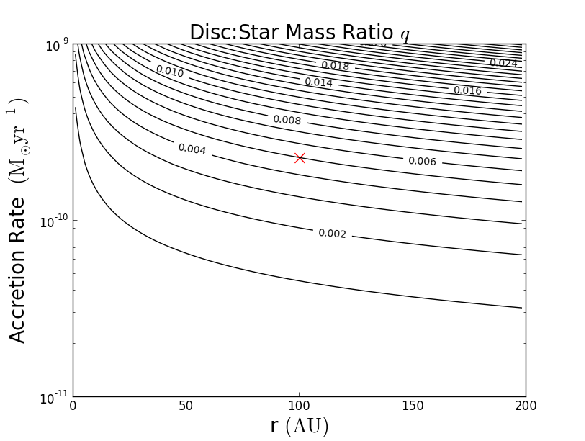}
\caption{Contour plot of disc-to-star mass ratio for accretion rate and radius, plotted in the range required for a host star of $2$ M$_{\odot}$ to have a quasi-steady, self gravitating disc of mass ratio $q=0.005$, or disc mass $M_D = 0.01$ M$_{\odot}$. The position of MWC 758 is marked by a red cross. In order for the system to be in a self-gravitating, quasi-steady state with a measured disc mass of $M_D = 0.01$ M$_{\odot}$ requires an accretion rate of $\sim 10^{-10}$ M$_{\odot}$ yr$^{-1}$.
\label{fig:MWC758contour}}
\end{figure}

We enter into our model a host star with mass 2 M$_{\odot}$, a disc outer radius of 100 AU and an accretion rate of $2\times 10^{-7}$ M$_{\mathrm{\odot}}$ yr$^{-1}$. In order for the disc to be in a quasi-steady, self-gravitating state for these specific parameters requires a disc-to-star mass ratio of $q\sim 0.25$. This gives a total disc mass that is over an order of magnitude larger than that given by \citet{andrewsetal2011}. This means that \emph{either}:

\begin{enumerate}[leftmargin=0.1in]
\item The spirals are due to self-gravity, and the mass of the disc surrounding MWC 758 has been underestimated by a factor of 50. Even if this is the case, \cite{donghallrice2015} have recently shown that self-gravitating spiral arms obey $m\sim 1/q$, suggesting that the expected dominant $m$-mode would be 4, \emph{if} the spirals are due to self-gravity. However, for $m=2$ spiral modes to dominate typically requires $q\gtrsim 0.5$, and that the accretion rate be high, of order $\sim \dot{M}\approx 10^{-6}$ M$_{\odot}$ yr$^{-1}$ \citep{donghallrice2015}. Such a disc would  have non-local angular momentum transport \citep{forganetal2011}, and as such would not be well-described by the viscous approximation in our analytical model.
\item The disc is self-gravitating, but the accretion rate is \emph{much} lower than any of the measured values given by \citet{andrewsetal2011} for MWC 758, and the measured disc mass is correct. Figure \ref{fig:MWC758contour} shows that for a disc around a host star of 2 M$_{\odot}$ to have a mass of $10^{-2}$ M$_{\odot}$ (or equivalently $q\sim 0.005$) requires that the accretion rate be of order $\sim 10^{-10}$ M$_{\odot}$ yr$^{-1}$. If this is the case, it is highly unlikely that spiral structure would be detectable since the $\alpha_{\mathrm{grav}}$, and therefore perturbation strength of the spiral, would be incredibly low.
\item Both the disc mass of MWC758 and the accretion rate are accurate. The spiral structure visible is due to some other mechanism, perhaps planet - disc interaction as discussed in \citet{benistyetal2015}, and not due to self-gravity. Accretion is therefore driven by something other than GI, such as MRI.
\end{enumerate}

Gravitational instabilities are certainly capable of producing structures which match the morphologies of observed low $m$-mode systems. However, it also makes demands on the system that in the case do not appear to be met, i.e. that disc-to-star mass ratio and accretion rate are very high.

Something else to bear in mind is that our analytic models make assumptions that are likely no longer valid in high mass ($q\gtrsim 0.5$) discs, in which global ($m\sim 2$) spiral modes dominate. When global torques are induced, the angular momentum transport is no longer local, and the semi-analytic models using a local viscous approximation are no longer justified. 

Additionally, this semi-analytic model uses the midplane cooling time to determine the effective gravitational $\alpha$. For massive discs, this will be largely underestimated compared to the actual $\alpha$ value in a global, radiatively cooling disc \citep{forganetal2011}. Given that the spirals in MWC 758 appear global in nature might suggest that we can't use our semi-analytic model in this comparison. However, producing such global spirals via GI would require disc properties even more discrepant than our model suggests, and so the basic conclusion would remain unchanged.

This should serve as a word of caution to the analysis of future observations of discs with non-axisymmetric structure. Modelling non-local discs in the local approximation will return discs with spiral amplitudes far lower than would realistically be present. On the other hand, such discs would be extremely massive and have high accretion rates. Not only is it unlikely that they would be confused for lower mass discs, they will also remain in this phase for a very short time. The local approximation is therefore probably reasonable for anything that is likely to be observed by ALMA.

Adding irradiation to MWC 758 would allow the system to maintain a larger disc mass. However, as shown in Section \ref{sec:results}, this removes spiral features from the disc. Adding just enough irradiation to the disc so that spiral features are still observable does not change the result we get for a system with parameters matching those of MWC 758. If the spirals present in MWC 758 are due to disc self-gravity, then they cannot be modelled using a local prescription of angular momentum transport, but this would also require the system to have very different properties to those observed.
\subsubsection{SAO 206462}
\label{subsubsec:SAO206462}
SAO 206462 is an isolated $1.7$ M$_{\odot}$ Herbig Ae/Be star at a distance of 142 pc in the constellation Lupus \citep{mulleretal2011}. It has a $\sim 10^{-3}$ M$_{\odot}$ disc \citep{thietal2001} and an accretion rate of $5.37\times 10^{-9}$ M$_{\odot}$ yr$^{-1}$ \citep{garcialopez2006} and an outer radius of 140 AU. Scattered light observations in NIR have revealed spiral structure in the outer disc \citep{mutoetal2012,garufietal2013}, and sub-mm ALMA observations have revealed large-scale asymmetries in the dust continuum \citep{perezetal2014}. These asymmetries have been fit using a vortex prescription, following \citet{regaly2012} by \citet{perezetal2014}, however, those authors concluded that the vortex prescription did not reproduce every observed feature, and significant residuals remained which coincided with the spiral arms seen in \emph{H}-band scattered light.

Although the disc mass is probably too low to trigger gravitational instabilities, as is the case with many T-Tauri stars there is evidence for systematic underestimation of the disc mass. We model SAO 206462 as if disc self-gravity is responsible for the spiral features present in the disc.  Using our model, to match the accretion rate of SAO 206462 requires a disc-to-star mass ratio of $q\sim 0.1$ in order for the disc to be in a quasi-steady, self gravitating state. 

Spiral arms which have GI as their origin make additional demands on a system that in this case do not seem to be fulfilled. The spirals must be compact (on scales less than $\sim 100$ AU), the disc must be massive ($q\gtrsim 0.25$) and the accretion rate must be high \citep{donghallrice2015}. This leaves us with the following available conclusions:
\begin{enumerate}[leftmargin=0.1in]%
\item The disc mass has been underestimated by several orders of magnitude, and the disc for SAO 206462 is actually well within the self-gravitating regime. However, even if this is the case, such a disc would not produce clear spiral structure due to the low $\alpha_{\mathrm{grav}}$ and therefore spiral amplitude. 
\item The accretion rate is much lower than that measured for SAO 206462, and the disc mass measured is correct. This would further decrease the amount of flux from the disc and again cause difficulty observing it. 
\item Both the disc mass and accretion rate are accurate, and the spiral features are not due to disc self-gravity.
\end{enumerate}
\subsubsection{HD 142527}
\label{subsubsec:HD142527}
The transition disc HD142527 has been observed in the near-IR, and has been revealed to have a unique morphology, appearing to consist of two bright arcs facing each other and one spiral arm \citep{fukagawa2006}. The central star's mass and age are respectively estimated at $1.9 - 2.2\mathrm{M}_{\odot}$ and $1-12\mathrm{Myr}$ \citep{fukagawa2006,verhoeff2011}. It has an accretion rate of $6.92 \times 10^{-8}\mathrm{M}_{\odot}$ $\mathrm{yr}^{-1}$ \citep{garcialopez2006} and the estimated flow rate of gas in the gap in the disc is between $7 \times 10^{-9}$ and $2 \times 10 ^{-7}\mathrm{M}_{\odot}$ $\mathrm{yr}^{-1}$ \citep{casassus2013}. The total disc mass has been measured from gas-to-dust ratios as $\sim$ 0.1 $\mathrm{M}_{\odot}$ \citep{verhoeff2011}. 

Spiral arms have been imaged in $^{12}$CO $J=2-1$ and $J=3-2$ using ALMA by \citet{codisc2014}, who placed lower limits on the mass of each spiral arm at $\sim 10^{-6}\mathrm{M_{\odot}}$. These features were interpreted as an acoustic wave launched by a planet (see e.g. \citealt{mutoetal2012}), however since it is now thought probable that HD 142527 has a low-mass stellar companion \citep{biller2012}, the spiral structures could certainly be tidally induced. GI is an alternative scenario which is able to replicate this grand design spiral structure, and since both \citet{codisc2014} and \citet{fukagawa2013} find $Q\sim 2.0$, there is evidence for gravitational instability being responsible for the spiral structure present.

We assume a 2 M$_{\odot}$ central star, and an accretion rate of $6.92 \times 10^{-8}\mathrm{M}_{\odot}$ $\mathrm{yr}^{-1}$. Since there is evidence that the disc may extend as far out as 600 AU in radius \citep{codisc2014}, we extend our disc out to 600 AU. 

To match the observed accretion rate of HD 142527 and extend out to 600 AU, such a disc would require an incredibly high disc-to-star mass ratio of $q\sim 0.75$. Such a disc would certainly have incredibly high global torques, and in reality would probably not survive in this quasi-steady, self-gravitating state. Therefore, the most likely explanation for the spiral structure observed in HD 142527 is not self-gravity.

An alternative explanation for the spiral structure is tidal interaction due to its low mass companion ($M_{\mathrm{companion}}\sim 0.1$ M$_{\odot}$), which is potentially on an eccentric orbit around HD 142527 \citep{fukagawa2006,bainesetal2006,biller2012,closeetal2014}.
\subsubsection{Conclusions from Observed Systems}
We examined the parameter space of three transition discs to determine if the non-axisymmetric structure which has been imaged in those discs could feasibly be due to disc self-gravity. For all three systems, it seems unlikely, unless the disc mass has been significantly underestimated. Even if the disc mass has been underestimated, and the disc is self-gravitating, we may expect to see a different number of $m$-modes dominant in the disc. Self-gravity imposes additional requirements on a system which do not seem to be consistent with the parameters of these systems.
\section{Discussion and Conclusion}
\label{sec:conclusions}
We performed an examination of the parameter space in which self-gravitating discs can exist, using a semi-analytical approach. We generated synthetic observations with the intention of investigating the range of accretion rates, disc masses and disc radii that would allow non-axisymmetric structure to be detected by ALMA. Our intention was not to reproduce the exact morphology of the observations, but rather to understand the strength of a perturbation required to generate an observable spiral arm.

Analytical models using a viscous prescription that assumes local angular momentum transport poorly describe systems which are dominant in the low $m$ spiral modes. Modelling non-local discs in the local approximation returns spiral amplitudes far lower than would be present in reality. If a quasi-steady, self-gravitating disc \emph{can} be described analytically using local transport, then there exists a small range of accretion rates for a given radius where the gravitational stress is high enough to generate observable spirals, but not so high as to cause the outer regions of the disc to fragment. However, non-local transport only becomes significant in discs with masses above half the mass of the central star, and such discs will probably have very short lives \citep{lodatorice2005,ricemayoarmitage2010}, so our analysis here is probably reasonable for anything that would be detected by ALMA.

Another important factor is external irradiation. If the accretion rate is close to the fragmentation limit, a small amount of external irradiation ($\sim$ 10 K) may prevent fragmentation with increasing accretion rate. If the accretion rate is well below the fragmentation limit, a small amount of irradiation ($\sim$ 10 K) causes the surface density profile of the outer part of the disc to be restructured, as the equilibrium disc structure is now more massive for a given accretion rate and radius. If infall from a natal cloud is occurring, this could well be a \emph{trigger} for fragmentation, as it is likely that in these regions the Jeans criterion would be satisfied. A moderate amount of irradiation ($\sim$ 30 K) can suppress fragmentation up to higher accretion rates, but at the cost of non-axisymmetric structure.

Ultimately, our results suggest that there is a relatively small range of parameter space in which a disc could be self-gravitating, not undergo fragmentation, and have spiral amplitudes large enough to be observable by ALMA. Broadly speaking, we would expect the disc mass to exceed $0.1$ M$_{\odot}$, the accretion rate to satisfy $10^{-7} \lesssim \dot{M} \lesssim 10^{-6}$ M$_{\odot}$ yr$^{-1}$ and the outer radius to be not much more than 100 AU. Additionally, the observing frequency and distance to the source also plays a role. We are more likely to observe spiral waves at 680 GHz than at 220 GHz, and it becomes increasingly difficult as the source distance increases.

 Although self gravitating discs can certainly match the morphology of observed systems, they also impose strict additional conditions which may not be met. In essence, our analysis suggests that there is a relatively small region of parameter space in which self-gravity may produce observable spiral features. We would therefore caution against interpreting such features as being due to disc self-gravity unless the disc is likely to fall into this region of parameter space.
\section{Acknowledgements}
We are very grateful to the anonymous referee, who's comments greatly improved the final version of this paper. CH warmly thanks Giovanni Dipierro and Guillaume Laibe for their insightful and elucidating discussion with her concerning dust grains in this model. KR and BB gratefully acknowledge support from STFC grant ST/M001229/1. DF gratefully acknowledges support from the ECOGAL ERC advanced grant. Some calculations for this paper were performed on the University of Exeter Supercomputer, a DiRAC Facility jointly funded by STFC, the Large Facilities Capital fund of BIS, and the University of Exeter, and on the Complexity DiRAC Facility jointly funded by STFC and the Large Facilities Capital Fund of BIS. TJH acknowledges funding from Exeter's STFC Consolidated Grant (ST/J001627/1).
\bibliographystyle{mn2e}
\bibliography{bib}{}

\begin{thebibliography}{70}
\expandafter\ifx\csname natexlab\endcsname\relax\def\natexlab#1{#1}\fi

\bibitem[{{Andrews} {et~al}\mbox{.}(2011){Andrews}, {Wilner}, {Espaillat},
  {Hughes}, {Dullemond}, {McClure}, {Qi}, \& {Brown}}]{andrewsetal2011}
{Andrews} S.~M., {Wilner} D.~J., {Espaillat} C., {Hughes} A.~M., {Dullemond}
  C.~P., {McClure} M.~K., {Qi} C., {Brown} J.~M., 2011, ApJ, 732, 42

\bibitem[{{Baines} {et~al}\mbox{.}(2006){Baines}, {Oudmaijer}, {Porter}, \&
  {Pozzo}}]{bainesetal2006}
{Baines} D., {Oudmaijer} R.~D., {Porter} J.~M., {Pozzo} M., 2006, MNRAS, 367,
  737

\bibitem[{{Bell} \& {Lin}(1994)}]{belllin1994}
{Bell} K.~R., {Lin} D.~N.~C., 1994, ApJ, 427, 987

\bibitem[{{Benisty} {et~al}\mbox{.}(2015){Benisty}, {Juhasz}, {Boccaletti},
  {Avenhaus}, {Milli}, {Thalmann}, {Dominik}, {Pinilla}, {Buenzli}, {Pohl},
  {Beuzit}, {Birnstiel}, {de Boer}, {Bonnefoy}, {Chauvin}, {Christiaens},
  {Garufi}, {Grady}, {Henning}, {Huelamo}, {Isella}, {Langlois}, {M{\'e}nard},
  {Mouillet}, {Olofsson}, {Pantin}, {Pinte}, \& {Pueyo}}]{benistyetal2015}
{Benisty} M. {et~al.}, 2015, A\&A, 578, L6

\bibitem[{{Biller} {et~al}\mbox{.}(2012){Biller}, {Lacour}, {Juh{\'a}sz},
  {Benisty}, {Chauvin}, {Olofsson}, {Pott}, {M{\"u}ller}, {Sicilia-Aguilar},
  {Bonnefoy}, {Tuthill}, {Thebault}, {Henning}, \& {Crida}}]{biller2012}
{Biller} B. {et~al.}, 2012, ApJL, 753, L38

\bibitem[{{Casassus} {et~al}\mbox{.}(2013){Casassus}, {van der Plas}, {M},
  {Dent}, {Fomalont}, {Hagelberg}, {Hales}, {Jord{\'a}n}, {Mawet},
  {M{\'e}nard}, {Wootten}, {Wilner}, {Hughes}, {Schreiber}, {Girard},
  {Ercolano}, {Canovas}, {Rom{\'a}n}, \& {Salinas}}]{casassus2013}
{Casassus} S. {et~al.}, 2013, Nature, 493, 191

\bibitem[{{Christiaens} {et~al}\mbox{.}(2014){Christiaens}, {Casassus},
  {Perez}, {van der Plas}, \& {M{\'e}nard}}]{codisc2014}
{Christiaens} V., {Casassus} S., {Perez} S., {van der Plas} G., {M{\'e}nard}
  F., 2014, ApJL, 785, L12

\bibitem[{{Clarke}(2009)}]{clarke2009}
{Clarke} C.~J., 2009, MNRAS, 396, 1066

\bibitem[{{Close} {et~al}\mbox{.}(2014){Close}, {Follette}, {Males}, {Puglisi},
  {Xompero}, {Apai}, {Najita}, {Weinberger}, {Morzinski}, {Rodigas}, {Hinz},
  {Bailey}, \& {Briguglio}}]{closeetal2014}
{Close} L.~M. {et~al.}, 2014, ApJL, 781, L30

\bibitem[{{Cossins}, {Lodato} \& {Clarke}(2009){Cossins}, {Lodato}, \&
  {Clarke}}]{cossinslodatoclarke2009}
{Cossins} P., {Lodato} G., {Clarke} C.~J., 2009, MNRAS, 393, 1157

\bibitem[{{Cossins}, {Lodato} \& {Testi}(2010){Cossins}, {Lodato}, \&
  {Testi}}]{cossinslodatotesti2010}
{Cossins} P., {Lodato} G., {Testi} L., 2010, MNRAS, 407, 181

\bibitem[{{Dipierro} {et~al}\mbox{.}(2014){Dipierro}, {Lodato}, {Testi}, \& {de
  Gregorio Monsalvo}}]{dipierro2014}
{Dipierro} G., {Lodato} G., {Testi} L., {de Gregorio Monsalvo} I., 2014, MNRAS,
  444, 1919

\bibitem[{{Dipierro} {et~al}\mbox{.}(2015){Dipierro}, {Pinilla}, {Lodato}, \&
  {Testi}}]{dipierro2015}
{Dipierro} G., {Pinilla} P., {Lodato} G., {Testi} L., 2015, MNRAS, 451, 974

\bibitem[{{Dong} {et~al}\mbox{.}(2015{\natexlab{a}}){Dong}, {Hall}, {Rice}, \&
  {Chiang}}]{donghallrice2015}
{Dong} R., {Hall} C., {Rice} K., {Chiang} E., 2015{\natexlab{a}}, ArXiv
  e-prints

\bibitem[{{Dong} {et~al}\mbox{.}(2015{\natexlab{b}}){Dong}, {Zhu}, {Rafikov},
  \& {Stone}}]{dongzhu2015}
{Dong} R., {Zhu} Z., {Rafikov} R.~R., {Stone} J.~M., 2015{\natexlab{b}}, ApJL,
  809, L5

\bibitem[{{Douglas} {et~al}\mbox{.}(2013){Douglas}, {Caselli}, {Ilee}, {Boley},
  {Hartquist}, {Durisen}, \& {Rawlings}}]{douglasetal2013}
{Douglas} T.~A., {Caselli} P., {Ilee} J.~D., {Boley} A.~C., {Hartquist} T.~W.,
  {Durisen} R.~H., {Rawlings} J.~M.~C., 2013, MNRAS, 433, 2064

\bibitem[{{Draine} \& {Lee}(1984)}]{drainelee}
{Draine} B.~T., {Lee} H.~M., 1984, ApJ, 285, 89

\bibitem[{{Dunham} {et~al}\mbox{.}(2014){Dunham}, {Stutz}, {Allen}, {Evans},
  {Fischer}, {Megeath}, {Myers}, {Offner}, {Poteet}, {Tobin}, \&
  {Vorobyov}}]{dunhametal2014}
{Dunham} M.~M. {et~al.}, 2014, ArXiv e-prints

\bibitem[{{Durisen} {et~al}\mbox{.}(2007){Durisen}, {Boss}, {Mayer}, {Nelson},
  {Quinn}, \& {Rice}}]{durisenetal2007}
{Durisen} R.~H., {Boss} A.~P., {Mayer} L., {Nelson} A.~F., {Quinn} T., {Rice}
  W.~K.~M., 2007, Protostars and Planets V, 607

\bibitem[{{Forgan} \& {Rice}(2011)}]{forganrice2011}
{Forgan} D., {Rice} K., 2011, MNRAS, 417, 1928

\bibitem[{{Forgan} \& {Rice}(2012)}]{forganrice2012}
{Forgan} D., {Rice} K., 2012, MNRAS, 420, 299

\bibitem[{{Forgan} \& {Rice}(2013{\natexlab{a}})}]{forganricejeans2013}
{Forgan} D., {Rice} K., 2013{\natexlab{a}}, MNRAS, 430, 2082

\bibitem[{{Forgan} \& {Rice}(2013{\natexlab{b}})}]{forganrice2013}
{Forgan} D., {Rice} K., 2013{\natexlab{b}}, MNRAS, 433, 1796

\bibitem[{{Forgan} {et~al}\mbox{.}(2011){Forgan}, {Rice}, {Cossins}, \&
  {Lodato}}]{forganetal2011}
{Forgan} D., {Rice} K., {Cossins} P., {Lodato} G., 2011, MNRAS, 410, 994

\bibitem[{{Fukagawa} {et~al}\mbox{.}(2006){Fukagawa}, {Tamura}, {Itoh}, {Kudo},
  {Imaeda}, {Oasa}, {Hayashi}, \& {Hayashi}}]{fukagawa2006}
{Fukagawa} M., {Tamura} M., {Itoh} Y., {Kudo} T., {Imaeda} Y., {Oasa} Y.,
  {Hayashi} S.~S., {Hayashi} M., 2006, ApJL, 636, L153

\bibitem[{{Fukagawa} {et~al}\mbox{.}(2013){Fukagawa}, {Tsukagoshi}, {Momose},
  {Saigo}, {Ohashi}, {Kitamura}, {Inutsuka}, {Muto}, {Nomura}, {Takeuchi},
  {Kobayashi}, {Hanawa}, {Akiyama}, {Honda}, {Fujiwara}, {Kataoka},
  {Takahashi}, \& {Shibai}}]{fukagawa2013}
{Fukagawa} M. {et~al.}, 2013, PASJ, 65, L14

\bibitem[{{Gammie}(2001)}]{gammie2001}
{Gammie} C.~F., 2001, ApJ, 553, 174

\bibitem[{{Garcia Lopez} {et~al}\mbox{.}(2006){Garcia Lopez}, {Natta}, {Testi},
  \& {Habart}}]{garcialopez2006}
{Garcia Lopez} R., {Natta} A., {Testi} L., {Habart} E., 2006, A\&A, 459, 837

\bibitem[{{Garufi} {et~al}\mbox{.}(2013){Garufi}, {Quanz}, {Avenhaus},
  {Buenzli}, {Dominik}, {Meru}, {Meyer}, {Pinilla}, {Schmid}, \&
  {Wolf}}]{garufietal2013}
{Garufi} A. {et~al.}, 2013, A\&A, 560, A105

\bibitem[{{Grady} {et~al}\mbox{.}(2013){Grady}, {Muto}, {Hashimoto},
  {Fukagawa}, {Currie}, {Biller}, {Thalmann}, {Sitko}, {Russell}, {Wisniewski},
  {Dong}, {Kwon}, {Sai}, {Hornbeck}, {Schneider}, {Hines}, {Moro
  Mart{\'{\i}}n}, {Feldt}, {Henning}, {Pott}, {Bonnefoy}, {Bouwman}, {Lacour},
  {Mueller}, {Juh{\'a}sz}, {Crida}, {Chauvin}, {Andrews}, {Wilner}, {Kraus},
  {Dahm}, {Robitaille}, {Jang-Condell}, {Abe}, {Akiyama}, {Brandner}, {Brandt},
  {Carson}, {Egner}, {Follette}, {Goto}, {Guyon}, {Hayano}, {Hayashi},
  {Hayashi}, {Hodapp}, {Ishii}, {Iye}, {Janson}, {Kandori}, {Knapp}, {Kudo},
  {Kusakabe}, {Kuzuhara}, {Mayama}, {McElwain}, {Matsuo}, {Miyama}, {Morino},
  {Nishimura}, {Pyo}, {Serabyn}, {Suto}, {Suzuki}, {Takami}, {Takato},
  {Terada}, {Tomono}, {Turner}, {Watanabe}, {Yamada}, {Takami}, {Usuda}, \&
  {Tamura}}]{gradyetal2013}
{Grady} C.~A. {et~al.}, 2013, ApJ, 762, 48

\bibitem[{{Harries} {et~al}\mbox{.}(2004){Harries}, {Monnier}, {Symington}, \&
  {Kurosawa}}]{harriesetal2004}
{Harries} T.~J., {Monnier} J.~D., {Symington} N.~H., {Kurosawa} R., 2004,
  MNRAS, 350, 565

\bibitem[{{Haworth} {et~al}\mbox{.}(2015){Haworth}, {Harries}, {Acreman}, \&
  {Bisbas}}]{haworthetal2015}
{Haworth} T.~J., {Harries} T.~J., {Acreman} D.~M., {Bisbas} T.~G., 2015, MNRAS,
  453, 2277

\bibitem[{{Hubeny}(1990)}]{hubeny1990}
{Hubeny} I., 1990, ApJ, 351, 632

\bibitem[{{Isella} {et~al}\mbox{.}(2010){Isella}, {Natta}, {Wilner},
  {Carpenter}, \& {Testi}}]{isellaetal2010}
{Isella} A., {Natta} A., {Wilner} D., {Carpenter} J.~M., {Testi} L., 2010, ApJ,
  725, 1735

\bibitem[{{Isella} {et~al}\mbox{.}(2008){Isella}, {Tatulli}, {Natta}, \&
  {Testi}}]{isellaetal2008}
{Isella} A., {Tatulli} E., {Natta} A., {Testi} L., 2008, A\&A, 483, L13

\bibitem[{{Kratter}, {Matzner} \& {Krumholz}(2008){Kratter}, {Matzner}, \&
  {Krumholz}}]{kratteretal2008}
{Kratter} K.~M., {Matzner} C.~D., {Krumholz} M.~R., 2008, ApJ, 681, 375

\bibitem[{{Kratter} {et~al}\mbox{.}(2010){Kratter}, {Matzner}, {Krumholz}, \&
  {Klein}}]{kratteretal2010a}
{Kratter} K.~M., {Matzner} C.~D., {Krumholz} M.~R., {Klein} R.~I., 2010, ApJ,
  708, 1585

\bibitem[{{Kratter} \& {Murray-Clay}(2011)}]{krattermurrayclay2011}
{Kratter} K.~M., {Murray-Clay} R.~A., 2011, ApJ, 740, 1

\bibitem[{{Kurosawa} {et~al}\mbox{.}(2004){Kurosawa}, {Harries}, {Bate}, \&
  {Symington}}]{kurosawaetal2004}
{Kurosawa} R., {Harries} T.~J., {Bate} M.~R., {Symington} N.~H., 2004, MNRAS,
  351, 1134

\bibitem[{{Laughlin} \& {Bodenheimer}(1994)}]{laughlin1994}
{Laughlin} G., {Bodenheimer} P., 1994, ApJ, 436, 335

\bibitem[{{Lodato} \& {Rice}(2004)}]{lodatorice2004}
{Lodato} G., {Rice} W.~K.~M., 2004, MNRAS, 351, 630

\bibitem[{{Lodato} \& {Rice}(2005)}]{lodatorice2005}
{Lodato} G., {Rice} W.~K.~M., 2005, MNRAS, 358, 1489

\bibitem[{{Lucy}(1999)}]{lucy1999}
{Lucy} L.~B., 1999, A\&A, 344, 282

\bibitem[{{Marino} {et~al}\mbox{.}(2015){Marino}, {Casassus}, {Perez}, {Lyra},
  {Roman}, {Avenhaus}, {Wright}, \& {Maddison}}]{marinoetal2015}
{Marino} S., {Casassus} S., {Perez} S., {Lyra} W., {Roman} P.~E., {Avenhaus}
  H., {Wright} C.~M., {Maddison} S.~T., 2015, ArXiv e-prints

\bibitem[{{Mathis}, {Rumpl} \& {Nordsieck}(1977){Mathis}, {Rumpl}, \&
  {Nordsieck}}]{grainsize}
{Mathis} J.~S., {Rumpl} W., {Nordsieck} K.~H., 1977, ApJ, 217, 425

\bibitem[{{McKee} \& {Ostriker}(2007)}]{mckeeostricker2007}
{McKee} C.~F., {Ostriker} E.~C., 2007, ARA\&A, 45, 565

\bibitem[{{McMullin} {et~al}\mbox{.}(2007){McMullin}, {Waters}, {Schiebel},
  {Young}, \& {Golap}}]{casa}
{McMullin} J.~P., {Waters} B., {Schiebel} D., {Young} W., {Golap} K., 2007, in
  Astronomical Society of the Pacific Conference Series, Vol. 376, Astronomical
  Data Analysis Software and Systems XVI, {Shaw} R.~A., {Hill} F., {Bell}
  D.~J., eds., p. 127

\bibitem[{{Meeus} {et~al}\mbox{.}(2012){Meeus}, {Montesinos},
  {Mendigut{\'{\i}}a}, {Kamp}, {Thi}, {Eiroa}, {Grady}, {Mathews}, {Sandell},
  {Martin-Za{\"i}di}, {Brittain}, {Dent}, {Howard}, {M{\'e}nard}, {Pinte},
  {Roberge}, {Vandenbussche}, \& {Williams}}]{meeusetal2012}
{Meeus} G. {et~al.}, 2012, A\&A, 544, A78

\bibitem[{{M{\"u}ller} {et~al}\mbox{.}(2011){M{\"u}ller}, {van den Ancker},
  {Launhardt}, {Pott}, {Fedele}, \& {Henning}}]{mulleretal2011}
{M{\"u}ller} A., {van den Ancker} M.~E., {Launhardt} R., {Pott} J.~U., {Fedele}
  D., {Henning} T., 2011, A\&A, 530, A85

\bibitem[{{Muto} {et~al}\mbox{.}(2012){Muto}, {Grady}, {Hashimoto}, {Fukagawa},
  {Hornbeck}, {Sitko}, {Russell}, {Werren}, {Cur{\'e}}, {Currie}, {Ohashi},
  {Okamoto}, {Momose}, {Honda}, {Inutsuka}, {Takeuchi}, {Dong}, {Abe},
  {Brandner}, {Brandt}, {Carson}, {Egner}, {Feldt}, {Fukue}, {Goto}, {Guyon},
  {Hayano}, {Hayashi}, {Hayashi}, {Henning}, {Hodapp}, {Ishii}, {Iye},
  {Janson}, {Kandori}, {Knapp}, {Kudo}, {Kusakabe}, {Kuzuhara}, {Matsuo},
  {Mayama}, {McElwain}, {Miyama}, {Morino}, {Moro-Martin}, {Nishimura}, {Pyo},
  {Serabyn}, {Suto}, {Suzuki}, {Takami}, {Takato}, {Terada}, {Thalmann},
  {Tomono}, {Turner}, {Watanabe}, {Wisniewski}, {Yamada}, {Takami}, {Usuda}, \&
  {Tamura}}]{mutoetal2012}
{Muto} T. {et~al.}, 2012, ApJL, 748, L22

\bibitem[{{Paczynski}(1978)}]{paczynski1978}
{Paczynski} B., 1978, Acta Astronomica, 28, 91

\bibitem[{{P{\'e}rez} {et~al}\mbox{.}(2014){P{\'e}rez}, {Isella}, {Carpenter},
  \& {Chandler}}]{perezetal2014}
{P{\'e}rez} L.~M., {Isella} A., {Carpenter} J.~M., {Chandler} C.~J., 2014,
  ApJL, 783, L13

\bibitem[{{Pinte} {et~al}\mbox{.}(2009){Pinte}, {Harries}, {Min}, {Watson},
  {Dullemond}, {Woitke}, {M{\'e}nard}, \& {Dur{\'a}n-Rojas}}]{pinteetal2009}
{Pinte} C., {Harries} T.~J., {Min} M., {Watson} A.~M., {Dullemond} C.~P.,
  {Woitke} P., {M{\'e}nard} F., {Dur{\'a}n-Rojas} M.~C., 2009, A\&A, 498, 967

\bibitem[{{Rafikov}(2005)}]{rafikov2005}
{Rafikov} R.~R., 2005, ApJL, 621, L69

\bibitem[{Regály {et~al}\mbox{.}(2012)Regály, Juhász, Sándor, \&
  Dullemond}]{regaly2012}
Regály Z., Juhász A., Sándor Z., Dullemond C.~P., 2012, Monthly Notices of
  the Royal Astronomical Society, 419, 1701

\bibitem[{{Rice} \& {Armitage}(2009)}]{ricearmitage2009}
{Rice} W.~K.~M., {Armitage} P.~J., 2009, MNRAS, 396, 2228

\bibitem[{{Rice} {et~al}\mbox{.}(2011){Rice}, {Armitage}, {Mamatsashvili},
  {Lodato}, \& {Clarke}}]{riceetal2011}
{Rice} W.~K.~M., {Armitage} P.~J., {Mamatsashvili} G.~R., {Lodato} G., {Clarke}
  C.~J., 2011, MNRAS, 418, 1356

\bibitem[{{Rice}, {Lodato} \& {Armitage}(2005){Rice}, {Lodato}, \&
  {Armitage}}]{rice2005}
{Rice} W.~K.~M., {Lodato} G., {Armitage} P.~J., 2005, MNRAS, 364, L56

\bibitem[{{Rice} {et~al}\mbox{.}(2004){Rice}, {Lodato}, {Pringle}, {Armitage},
  \& {Bonnell}}]{riceetal2004}
{Rice} W.~K.~M., {Lodato} G., {Pringle} J.~E., {Armitage} P.~J., {Bonnell}
  I.~A., 2004, MNRAS, 355, 543

\bibitem[{{Rice}, {Mayo} \& {Armitage}(2010){Rice}, {Mayo}, \&
  {Armitage}}]{ricemayoarmitage2010}
{Rice} W.~K.~M., {Mayo} J.~H., {Armitage} P.~J., 2010, MNRAS, 402, 1740

\bibitem[{{Ruge} {et~al}\mbox{.}(2013){Ruge}, {Wolf}, {Uribe}, \&
  {Klahr}}]{rugeetal2013}
{Ruge} J.~P., {Wolf} S., {Uribe} A.~L., {Klahr} H.~H., 2013, A\&A, 549, A97

\bibitem[{{Shakura} \& {Sunyaev}(1973)}]{shakurasunyaev1973}
{Shakura} N.~I., {Sunyaev} R.~A., 1973, A\&A, 24, 337

\bibitem[{{Spitzer}(1942)}]{spitzer1942}
{Spitzer}, Jr. L., 1942, ApJ, 95, 329

\bibitem[{{Terebey}, {Shu} \& {Cassen}(1984){Terebey}, {Shu}, \&
  {Cassen}}]{terebey1984}
{Terebey} S., {Shu} F.~H., {Cassen} P., 1984, ApJ, 286, 529

\bibitem[{{Thi} {et~al}\mbox{.}(2001){Thi}, {van Dishoeck}, {Blake}, {van
  Zadelhoff}, {Horn}, {Becklin}, {Mannings}, {Sargent}, {van den Ancker},
  {Natta}, \& {Kessler}}]{thietal2001}
{Thi} W.~F. {et~al.}, 2001, ApJ, 561, 1074

\bibitem[{{Toomre}(1964)}]{toomre1964}
{Toomre} A., 1964, ApJ, 139, 1217

\bibitem[{{van Leeuwen}(2007)}]{vanleeuwen2007}
{van Leeuwen} F., 2007, A\&A, 474, 653

\bibitem[{{Verhoeff} {et~al}\mbox{.}(2011){Verhoeff}, {Min}, {Pantin},
  {Waters}, {Tielens}, {Honda}, {Fujiwara}, {Bouwman}, {van Boekel},
  {Dougherty}, {de Koter}, {Dominik}, \& {Mulders}}]{verhoeff2011}
{Verhoeff} A.~P. {et~al.}, 2011, A\&A, 528, A91

\bibitem[{{Vorobyov} \& {Basu}(2010)}]{vorobyovbasu2010}
{Vorobyov} E.~I., {Basu} S., 2010, ApJL, 714, L133

\bibitem[{{Whitworth} {et~al}\mbox{.}(2010){Whitworth}, {Stamatellos}, {Walch},
  {Kaplan}, {Goodwin}, {Hubber}, \& {Parker}}]{whitworthetal2010}
{Whitworth} A., {Stamatellos} D., {Walch} S., {Kaplan} M., {Goodwin} S.,
  {Hubber} D., {Parker} R., 2010, in IAU Symposium, Vol. 266, IAU Symposium,
  {de Grijs} R., {L{\'e}pine} J.~R.~D., eds., pp. 264--271

\end{thebibliography}


\label{lastpage}

\end{document}